\documentclass[graybox]{svmult}

\usepackage{mathptmx}
\usepackage{amssymb}
\usepackage{amsmath}
\usepackage{amsfonts}
\usepackage{helvet}         
\usepackage{courier}        
\usepackage{type1cm}        
\usepackage{makeidx}         
\usepackage{graphicx}        
\usepackage{multicol}        
\usepackage[bottom]{footmisc}

\makeindex             

\begin{document}

\title*{Fluctuations, dissipation and the dynamical Casimir effect}

\author{Diego A. R. Dalvit, Paulo A. Maia Neto, and Francisco Diego Mazzitelli}
\institute{
 Diego A. R. Dalvit  \at Theoretical Division MS B213, Los Alamos National Laboratory, Los Alamos, NM 87545, USA\\ \email{dalvit@lanl.gov}
\and Paulo A. Maia Neto \at Instituto de F\'{\i}sica UFRJ, Caixa Postal 68528, Rio de Janeiro RJ  21941-972, Brazil\\ \email{pamn@if.ufrj.br}
\and
Francisco Diego Mazzitelli \at  
Departamento de F\'{\i}sica, FCEyN, Universidad de Buenos Aires and IFIBA, CONICET,
Ciudad Universitaria, Pabell\'on I, 1428 Buenos Aires, Argentina\\ \email{fmazzi@df.uba.ar}
}
%
\maketitle

\abstract{Vacuum fluctuations provide a fundamental source of dissipation for
systems coupled to quantum fields by radiation pressure. In the
dynamical Casimir effect, accelerating neutral bodies in free space
give rise to the emission of real photons while experiencing a damping
force which plays the role of a radiation reaction force. Analog
models where non-stationary conditions for the electromagnetic field
simulate the presence of moving plates are currently under
experimental investigation. A dissipative force might also appear in
the case of uniform relative motion between two bodies, thus leading
to a new kind of friction mechanism without mechanical contact.  In
this paper, we review recent advances on the dynamical Casimir and
non-contact friction effects, highlighting their common physical
origin.}

\section{Introduction}
\label{sec:intro}

The Casimir force discussed  in this volume  represents the average 
radiation pressure force upon one of the interacting bodies.
When the zero-temperature limit is considered, the average is taken over
the  vacuum field state. Although the average electric and magnetic fields vanish, the Casimir force is finite because  radiation pressure is
quadratic in the field strength operators. In this sense, the Casimir force derives from the fluctuating fields associated with the 
field zero-point  energy (or more precisely from their modification by the interacting bodies).
 
As any quantum observable, the radiation pressure itself fluctuates \cite{Barton91A,Barton91B}. 
For a single body at rest in empty space, the average vacuum radiation pressure  vanishes (for the ground-state field cannot be an energy source), and all that 
is left is a fluctuating force driving a quantum Brownian motion \cite{Jaekel93}. 
The resulting dynamics is characterized by diffusion in phase space, thus leading to decoherence of the body center-of-mass \cite{Dalvit00}. 

Besides diffusion, the radiation pressure coupling also leads to dissipation, with the corresponding coefficients connected by the
fluctuation-dissipation theorem \cite{Callen51}. 
As in the classical Brownian motion, the fluctuating force on the body at rest  is closely related to a dissipative force exerted when the body is set in motion.
Since the vacuum state is Lorentz invariant, the Casimir dissipative force vanishes in the case of uniform motion of a single body in 
empty space,
as expected from the principle of relativity.
For a non-relativistic ``mirror'' in one spatial dimension (1D), the Casimir dissipative force is proportional to the  
second-order derivative of the velocity \cite{Ford82}, 
like the radiation reaction force in classical electrodynamics. 
Casimir dissipation is in fact connected to the emission of photon pairs by the accelerated (electrically neutral) mirror,
an effect known  as 
the dynamical Casimir effect (DCE). 
The power dissipated in the motion of the mirror is indeed equal to the total radiated power in DCE as expected from energy conservation. 

The creation of photons in  a 1D cavity with one moving mirror was first analyzed by Moore \cite{Moore70}, and  explicit results were later 
derived in Ref. \cite{Castagnino84}.  Relativistic results for the dissipative Casimir force upon a single mirror in 1D 
and the connection with DCE were derived in a seminal paper 
by Fulling and Davies \cite{Fulling76}. 
At this earlier stage, the main motivation  was the analogy with the Hawking radiation associated with black-hole evaporation \cite{Hawking75}. 

The interplay between  Casimir dissipation and  fluctuations was  investigated only much later  \cite{Jaekel93,Braginsky91,Jaekel92A}, 
in connection  with  a major issue in quantum optics:
the fundamental quantum limits of position measurement (this  was  motivated by  the quest for
interferometric detection of gravitational waves) \cite{quantum_limit}. 
Linear response theory \cite{Kubo} provides 
a valuable tool for computing the Casimir dissipative force on a moving body from the fluctuations of the force on the body at rest, which is in general much simpler to 
calculate. This method was employed to compute the dissipative force on a 
moving, perfectly reflecting sphere \cite{Neto93} and on a 
plane surface experiencing a 
time-dependent perturbation \cite{Barton93book}
(the latter was also computed by taking the full time-dependent boundary
conditions (b.c.) \cite{Neto95,Golestanian98}). 

In all these configurations, Casimir dissipation turns out to be very small when 
realistic physical parameters are taken into account. 
The predicted orders of magnitude are more promising when considering a closely related effect: quantum
non-contact friction in the shear relative motion between two parallel surfaces \cite{Volotikin2007,Pendry1997}.
In contrast with  the radiation reaction dissipative effect discussed so far, quantum friction takes place even for a uniform relative motion. 
On the other hand, quantum friction
requires the material media to have  finite response times (dispersion). From Kramers-Kronig relations \cite{Nussenzveig}, the material media must
also be dissipative, and the resulting friction depends on the imaginary part of the dielectric constant $\epsilon.$

Whereas the direct measurement of the  Casimir radiation reaction dissipative force seems to be beyond hope,
the corresponding photon emission effect might be within reach in the near future. The properties of the radiated photons have been 
analyzed in great detail over recent years. 
For a single moving  plane mirror, the frequency spectrum was computed in 1D 
\cite{Lambrecht96}
as well as in 3D \cite{Neto96} in the non-relativistic approximation. A variety of 3D geometries were considered, including
deforming mirrors \cite{Montazeri08},
parallel plates  \cite{Mundarain98}, 
cylindrical waveguides \cite{Neto05}, 
and spherical cavities containing either scalar \cite{Pascoal09} or  
electromagnetic fields \cite{Eberlein96, Mazzitelli06}.

Closed rectangular \cite{Dodonov96, Crocce01, Crocce02} or cylindrical \cite{Crocce05} microwave cavities with one moving wall are by far
the best candidates for a possible experimental implementation, with the mechanical oscillation frequency $\Omega$  tuned into parametric resonance 
with cavity field modes. 
Because of the parametric amplification effect, it is necessary to go beyond the perturbative approximation in order to compute the intracavity 
photon number even in the non-relativistic regime \cite{Dalvit98, Lambrecht98, Dalvit99}.

As the microwave field builds up inside the cavity, cavity losses (due to transmission, dissipation or diffraction at the rough 
cavity walls) become increasingly important. 
Finite transmission at the mirrors of a 1D cavity was taken into account
within the scattering approach developed in Refs.~\cite{Lambrecht96, Jaekel92C,Dezael10}. 
Master equations for the  reduced density operator of the cavity field  in lossy 3D cavities were derived in Refs. \cite{Dodonov98,Schaller02}.
Predictions for the total photon number produced at very long times obtained from 
the different models are in 
disagreement, so that  
a reliable estimation of the DCE magnitude under realistic conditions is still an
open theoretical problem. 

It is nevertheless clear that measuring DCE photons is  a highly non-trivial challenge (see for instance the proposal \cite{Kim06}
based on superradiance amplification).  For this reason,   in recent years the focus has been re-oriented towards analog models of DCE. 
Although dynamical Casimir photons are in principle emitted even in the case of a global `center-of-mass' oscillation of a cavity, the orders of magnitude are 
clearly more favorable when some cavity length is modulated. In this case, 
one might  modulate the optical cavity length by 
changing the intracavity refractive index (or more generally  material optical constants) instead of changing the physical cavity length. 
For instance, 
the conductivity of a semiconductor slab can be rapidly changed with the help of a short optical pulse, simulating the motion of a metallic mirror 
\cite{Yablonovitch89, Lozovik95}
and thereby producing photons exactly as in the DCE \cite{Crocce04,Mendonca05}.
An experiment along these lines is currently under way \cite{Braggio05} (see Ref. \cite{Braggio09} for an update). 
Alternatively, one might 
select a setup for
operation of an optical parametric oscillator  such that it becomes formally equivalent to DCE with a modulation frequency in the 
optical domain \cite{Dezael10}. 

Alongside the examples in quantum and nonlinear optics, 
one can also devise additional  analogues of DCE in the field of circuit quantum electrodynamics \cite{circuitQED}. For instance, a co-planar waveguide with a 
superconducting quantum interference device
(SQUID) 
at its end is formally equivalent to a 1D model for a single  mirror \cite{Nori09,Delsing10}.
When a time-dependent magnetic flux is applied to the SQUID, it simulates the motion of the mirror. 
More generally, Bose-Einstein condensates also provide interesting analogues for DCE \cite{Carusotto09} and Casimir-like dissipation \cite{Roberts}, 
with  electromagnetic vacuum fluctuations replaced by zero-point fluctuations of the condensate. 

Reviews on fluctuations and Casimir dissipation on one hand and on 
DCE on the other hand
can be found in  Refs. \cite{Jaekel97, Golestanian99}
and Refs. \cite{Dodonov01,Dodonov09b,Dodonov10}, respectively. 
This  review paper  is organized as follows. In Sec. 2, we discuss the interplay between fluctuations, dissipation and the photon creation effect 
for  a single mirror in free space. Sec. 3 presents a short introduction to non-contact quantum friction. In Sec. 4, 
photon creation in 
resonant cavities with either moving walls or time-dependent material properties is  presented in detail. Sec. 5 briefly discusses experimental 
proposals, and Sec. 6 contains some final remarks. 

\section{Dissipative effects of  the quantum vacuum}

\subsection{1D models}

We start with the simplest theoretical model: a non-relativistic point-like  `mirror'  coupled to a massless scalar field $\phi(x,t)$ in 1D. We take the Dirichlet b.c. at the 
instantaneous mirror position $q(t):$
\begin{equation}
\label{Dirichlet}
\phi(q(t),t)=0.
\end{equation}
In the non-relativistic approximation, we expect the vacuum radiation pressure force $f(t)$
to be proportional to some derivative of the mirror's velocity.
As a quantum effect, the force 
must also be proportional to $\hbar,$ and then dimensional analysis yields
\begin{equation}
\label{F1+1}
f(t) \propto \frac{\hbar  q^{(3)}(t)}{c^2}, \hspace{30pt} (1{\rm D})
\end{equation}
where $q^{(n)}(t)\equiv d^{n}q(t)/dt^n.$ Note that (\ref{F1+1}) is consistent with the Lorentz invariance of the vacuum field state, which excludes
friction-like forces proportional to $q^{(1)}(t)$ for a single moving mirror (but not in the case of relative motion between two mirrors discussed in the next two sections). 

In order to  compute the dimensionless prefactor in (\ref{F1+1}),
we solve the b.c. (\ref{Dirichlet})  to first order in $q(t)$ as in Ref.~\cite{Ford82}, with the mirror's motion treated as a small perturbation.
However, instead of 
analyzing in the time domain, we  switch to the frequency domain, 
which allows us to understand more clearly the region of validity of  the  theoretical model 
leading to (\ref{F1+1}). We write the Fourier transform of the field as a perturbative expansion:
\begin{equation}
\label{expansion}
\Phi(x,\omega)= \Phi_0(x,\omega)+\delta\Phi(x,\omega),
\end{equation}
where the unperturbed field $\Phi_0(x,\omega)$ corresponds to a static mirror at $x=0:$
\(
\Phi_0(0,\omega)\equiv 0.  
\)
The boundary condition for $\delta\Phi(x,\omega)$ is derived from (\ref{Dirichlet}) by taking a Taylor expansion around $x=0$ to first order in $Q(\Omega)$ 
(Fourier transform of $q(t)$):
\begin{equation}
\label{sideband}
\delta\Phi(0,\omega_{\rm o}) = -\int_{-\infty}^{\infty}\frac{d\omega_{\rm i}}{2\pi} \, Q(\omega_{\rm o}-\omega_{\rm i}) \,\partial_x\Phi_0(0,\omega_{\rm i}).
\end{equation}
Eq.~(\ref{sideband}) already contains the frequency modulation effect at the origin of Casimir dissipation: the motion of the mirror 
(frequency $\Omega$) generates  an output   amplitude at the sideband frequency $\omega_{\rm o}= \omega_{\rm i}+\Omega$ 
proportional to $Q(\Omega)$ from a given  input field frequency $\omega_{\rm i}.$ 

In order to find the Casimir dissipative force,
we take the Fourier transform of the appropriate component $T_{11}= \frac{1}{2} [\frac{1}{c^2}(\partial_t\phi)^2+(\partial_x \phi)^2]$
of the energy-momentum tensor and then replace the total field $\Phi(x,\omega)$ containing the solution 
$\delta\Phi(x,\omega)$ of eq.~(\ref{sideband}). After 
averaging over the vacuum state, the resulting  force is written as 
\begin{eqnarray}
F(\Omega)& = &  \chi(\Omega) Q(\Omega) , \\
\label{chi1}
\chi(\Omega) & = &2i\frac{\hbar}{c^2} \int_{-\infty}^{\infty}\frac{d\omega_{\rm i}}{2\pi} (\Omega+\omega_{\rm i}) |\omega_{\rm i}| .
\end{eqnarray}

After regularization, it is simple to show that the contribution $\int_{-\infty}^{-\Omega}d\omega_{\rm i}(...)$ cancels the contribution  $\int_{0}^{\infty}d\omega_{\rm i}(...)$ in (\ref{chi1}).
Thus, only field frequencies  in the interval $-\Omega\le  \omega_{\rm i}\le 0$ contribute to the dynamical radiation pressure force, yielding
\begin{equation}
\label{final}
F(\Omega) = i\frac{\hbar\Omega^3}{6\pi c^2} Q(\Omega) ,
\end{equation}
in agreement with (\ref{F1+1}) with a positive prefactor ($\frac{1}{6\pi}$) as expected for a dissipative force.  
This result was first obtained in Ref.~\cite{Ford82} within the perturbative approach and coincides with the non-relativistic limit 
of the exact result derived much earlier in Ref.~\cite{Fulling76}. It may also be obtained as a limiting case of the result for a partially transmitting mirror, which was 
derived either from the perturbative approach outlined here \cite{Jaekel92A} or by developing the appropriate Schwinger-Keldysh effective action within a functional approach to the dissipative Casimir effect \cite{Fosco07}. 
When considering a moving dielectric half-space, one obtains as expected one-half of the r.h.s. of (\ref{final})  in the 
limit of an infinite refractive index  \cite{Barton93}. The final result for $F(\Omega)$ is exactly the same as in (\ref{final}) when
we replace the Dirichlet b.c. (\ref{Dirichlet}) by the 
Neumann b.c.
at the instantaneously co-moving Lorentz frame (primed quantities refer to the co-moving frame)
$\partial_{x'} \phi|_{x'=q(t')}=0$ 
\cite{Alves03}.
Dirichlet and Neumann b.c.  also yield  the same force in the more general relativistic regime \cite{Alves08}. 
On the other hand,  for the Robin b.c. 
\begin{equation}
\label{Robin}
\partial_{x'}\phi|_{x'=q(t')}=\frac{1}{\beta}\phi|_{x'=q(t')} 
\end{equation}
($\beta$ is a constant parameter), 
the force susceptibility $\chi(\Omega)$
displays a non-monotonic dependence on $\Omega,$ and is nearly suppressed at
$\Omega\sim 2.5\, c/\beta$ \cite{Mintz06}.  

\begin{figure}[b]
\begin{center}
\includegraphics[scale=.65]{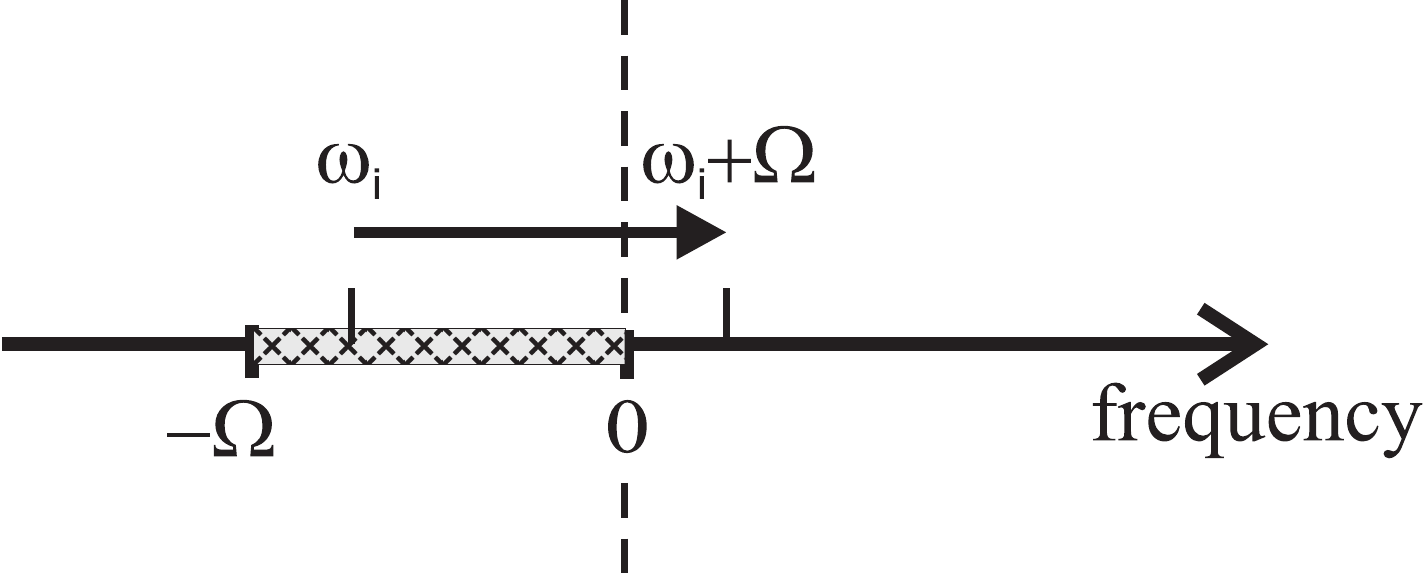}
\end{center}
\caption{For a given mechanical modulation frequency $\Omega$ (assumed to be positive in this diagram), the field modes contributing to the dissipative dynamical Casimir effect lie in the  interval $[-\Omega,0],$ which corresponds to negative input field frequencies $\omega_{\rm i}$, yielding positive sideband frequencies.  }
\label{fig_sideband}      
\end{figure}

We could have anticipated that 
high-frequency modes with $\omega_{\rm i}\gg \Omega$ would not contribute because they ``see''  the mirror's motion at frequency $\Omega$ as a quasi-static perturbation, and indeed the dissipative Casimir force  originates from low-frequency modes for which the motion is non-adiabatic. 
But there is a more illuminating interpretation that explains why the contribution comes precisely from the frequency interval $[-\Omega,0].$ 
In Fig.~\ref{fig_sideband}, we show that this interval
corresponds to the field modes leading to frequency sidebands $\omega_{\rm o}=\omega_{\rm i}+\Omega$ (see eq. (\ref{sideband})) 
{\it across} the border between positive and negative frequencies (for $\Omega<0$ the corresponding  interval is $[0,-\Omega]$ and the analysis is essentially
the same). 
In other words, for these specific modes the motional frequency modulation leads to mixing between positive and negative frequencies.
Bearing in mind the correspondence between positive (negative) frequencies and annihilation (creation) operators, 
this mixing translates into a Bogoliubov transformation coupling output annihilation operators  to input creation ones, and viceversa \cite{Dodonov90}
(examples will be presented in Section 4). 
The important conclusion is that
sideband generation 
for these modes   corresponds to photon creation (and also annihilation in the case of a general initial field state), whereas  
outside the interval $[-\Omega,0]$, where no mixing occurs,   the sideband effect corresponds to photon inelastic scattering with neither  creation nor annihilation. 
 
From this discussion, we can also surmise the important property that the dynamical Casimir photons have frequencies bounded by the mechanical frequency $\Omega,$ 
as long as the perturbative non-relativistic approximation holds \cite{Lambrecht96}. Moreover, the Casimir photons are emitted in pairs, with photon 
frequencies satisfying $\omega_1+\omega_2=\Omega.$ Hence the frequency spectrum is symmetrical around $\Omega/2$, where it has a maximum 
for the Dirichlet case  \cite{Lambrecht96} but not generally for the Robin b.c. \cite{Mintz06B}. 
 
In short, the derivation of (\ref{F1+1}) from (\ref{sideband}) 
highlights the direct connection between the dissipative dynamical  force and the 
dynamical Casimir photon emission effect. The dissipative force thus plays the role of a Casimir radiation reaction force, damping the 
mirror's mechanical energy as Casimir photons are emitted. In fact,  
the expression given by (\ref{F1+1}) has the same form as 
the radiation reaction force in classical electrodynamics, apart from a dimensionless pre-factor inversely proportional 
to the fine-structure constant $e^2/(\hbar c).$ 

The discussion on the field modes actually contributing in (\ref{sideband})   also allows us to address the domain of validity  of the various assumptions employed in the derivation  presented above.  We
have assumed in (\ref{Dirichlet}) that the field vanishes at the instantaneous mirror's position, no matter how fast the mirror and field oscillate. 
However, the electric currents and charge density inside a real metallic mirror respond to field and position changes over a finite time scale, so that we expect our 
oversimplified model to be physically meaningful only at low frequencies (more general results and discussions are presented in Refs.~\cite{Jaekel92A, Jaekel92B}). 
Since the relevant field frequencies are bounded by the mechanical frequencies ($|\omega_{\rm i}| \le |\Omega|$), 
the model is consistent as long as typical mechanical frequencies are much smaller than the frequency scales characterizing the metallic response - typically the plasma frequency of metals. 

A second point to be clarified is the connection between the perturbative linear approximation employed above and the non-relativistivistic approximation.
When deriving (\ref{sideband}) from (\ref{Dirichlet}), we have taken the long-wavelength approximation to expand the field around $x=0.$ 
Let us assume, to simplify the discussion, that the mirror oscillates with frequency $\Omega$ and amplitude $q_0.$ The non-relativistic regime then translates into
$\Omega q_0/c\ll 1$
and all relevant field modes correspond to long wavelengths $2\pi c/\omega_{\rm i}\gg q_0$
since they 
satisfy the inequality $|\omega_{\rm i}| \le \Omega.$ 
More generally, the long-wavelength approximation follows from the non-relativistic condition provided that
there is an inertial reference frame for which the motion is spatially bounded.

\subsection{Casimir-driven decoherence}

As in classical Brownian motion, the dissipative effect  is closely related to the fluctuations provided by the 
reservoir, which in our case is the quantum vacuum field. 
This connection  
provides yet  another  tool for computing  the dissipative Casimir force:   by using linear response theory, 
one can derive  (\ref{final})
from the correlation function of the force on a static plate \cite{Jaekel92A}.
(This method has been employed for different geometries in 3D \cite{Braginsky91, Neto93, Barton93book, Neto05}). 
More interestingly, if we take the mirror's position as a full  dynamical observable rather than a prescribed function of time, 
vacuum radiation pressure plays the role of a  Langevin fluctuating force \cite{Jaekel93, Fosco07}, leading to diffusion in 
phase space, which adds to the associated average dissipative force (\ref{final}).

Let us first analyze the 
mechanical effect of the dissipative Casimir force in the context of 
classical dynamics.
We consider a point-like mirror of mass $M$ in a harmonic potential well of frequency $\Omega.$ 
Taking the Casimir dissipative force given by (\ref{final}) into account, and neglecting for the time being any associated stochastic force, we write the 
mirror's equation of motion as 
\begin{equation}
\label{eq_motion}
\frac{d^2q(t)}{dt^2}= - \Omega^2 q(t) +\frac{\hbar}{6\pi M c^2} \,\frac{d^3q(t)}{dt^3}.
\end{equation}
We assume that the oscillator's zero-point energy $\hbar \Omega/2$ is much smaller than the rest mass energy
$M c^2,$ and then find oscillatory solutions of (\ref{eq_motion}) which are damped at the rate 
\begin{equation}
\label{Gamma}
{\Gamma} = \frac{1}{12\pi}\,\frac{\hbar \Omega}{ Mc^2}\,\Omega   \ll \Omega .
\end{equation}
This result provides a good illustration of how weak the Casimir dissipation is for a single mirror in a vacuum. On the other hand, the associated diffusion 
in phase space is more relevant, particularly in the context of the full quantum theory discussed below, since it provides an efficient  decoherence mechanism for 
non-classical quantum states. 

The quantum description of the mirror's dynamics can be developed from 
the Hamiltonian for the radiation pressure coupling with a 
dispersive semi-transparent mirror (transparency frequency $\omega_c$) \cite{Barton95}. 
One  derives a master equation for the reduced density operator $\rho$ of the mirror, which 
can also be cast in the form of a Fokker-Planck equation for the Wigner function $W(x,p,t)$ representing the mirror quantum state \cite{Dalvit00}:
\begin{equation}
\label{Fokker}
\partial_tW= -(1-\Delta M/M)\frac{p}{M}\partial_xW + M\Omega^2 x\partial_p W + 2\Gamma \partial_x(xW) +D_1\frac{\partial^2W}{\partial x^2}  - D_2 \frac{\partial^2W}{\partial x\partial p} .
\end{equation}
The time-dependent coefficients $\Delta M$ (mass correction), $\Gamma$ (damping coefficient), $D_1$ and $D_2$ (diffusion coefficients) are written in terms of the correlation function of the field 
linear momentum operator. The perfectly reflecting limit corresponding to (\ref{Dirichlet}) is obtained when $\Omega\ll \omega_c,$ in line with our previous discussion since 
$1/\omega_c$ represents the characteristic  time scale for the material medium response. In this limit, for $t\gtrsim 1/\omega_c$, 
$\Gamma(t)$ rapidly approaches the expected constant value as given 
by (\ref{Gamma}), whereas $D_1(t)$ approaches the asymptotic value 
\begin{equation}
\label{D1_Gamma}
D_1= \hbar \Gamma/(M\Omega) 
\end{equation}
for $t\gtrsim 2\pi/\Omega.$ This connection between diffusion and damping plays the role of a fluctuation-dissipation theorem for the vacuum state (zero temperature). 
 
Under the time evolution described by the Fokker-Planck equation (\ref{Fokker}), an initially pure state gradually evolves into a statistical mixture. The 
physical reason behind this decoherence effect is  the buildup of entanglement between the mirror and  
the field due to the radiation pressure coupling and the associated dynamical Casimir photon creation \cite{Dalvit00}. 
As an example of initial state, we consider the ``Schr\"odinger's cat'' superposition of coherent states
\(
|\psi\rangle = (|\alpha\rangle + |-\alpha\rangle)/\sqrt{2}
\)
with the amplitude $\alpha=iP_0/\sqrt{2M\hbar\omega_0}$ along the imaginary axis
($\pm P_0$ are the average momenta corresponding to each state's component).
The corresponding Wigner function  $W(x,p)$ contains an oscillating term proportional to
$\cos(2P_0 x/\hbar)$ which represents the coherence of the superposition.  Clearly, the diffusion term proportional to $D_1$ in 
(\ref{Fokker}) will wash out these oscillations along the position axis, thus transforming the cat state into a mixture of the two coherent states.
From (\ref{Fokker}) and (\ref{D1_Gamma}), the corresponding decoherence time scale is 
\begin{equation}
t_d = \frac{\hbar^2}{2P_0^2 D_1}=4 \left(\frac{\delta p}{2P_0}\right)^2 \, \Gamma^{-1} ,
\label{td}
\end{equation}
where 
$\delta p= \sqrt{M\hbar \Omega/2}$ is the momentum uncertainty of the coherent state (we have assumed that $\Omega t_d\gg 1$). 
Since $2P_0$ measures the distance between the two components in phase space, (\ref{td}) shows that decoherence is stronger 
when the state components are further apart, corresponding to  thinner interference fringes in phase space. 
Decoherence from entanglement driven by dynamical Casimir photon creation is thus very 
effective for macroscopic superpositions in spite of the smallness of the corresponding damping coefficient $\Gamma.$ As the radiation pressure 
control of  microresonators 
improves all the way  to the quantum level \cite{microresonators}, Casimir driven decoherence might eventually become of experimental relevance. 
 
\subsection{3D models}
 
The orders of magnitude can  be reliably  assessed  only by considering the real-world three-dimensional space. 
We start with the simplest geometry in 3D: a plane  mirror parallel to the $xy$ plane, of area $A$ and moving along the $z-$axis. 
For an infinite plane, we expect the dissipative Casimir force to be proportional to $A,$ so that we have to modify (\ref{F1+1}) to include a squared length.
For a scalar field satisfying a Dirichlet b.c. analogous to (\ref{Dirichlet}), the
derivation is very similar to the one outlined above
\cite{Ford82}:
\begin{equation}
\label{3scalar}
f(t) = -\frac{\hbar A  q^{(5)}(t)}{360 \pi^2 c^4}. \hspace{30pt} (3\mbox{D,  scalar})
\end{equation}

For the electromagnetic field, the 
model of perfect reflectivity provides an accurate description of metallic mirrors at low frequencies. For a  mirror moving along the $z$-axis,  the 
electric and magnetic  fields  in the instantaneously co-moving Lorentz frame $S'$ satisfy
\begin{equation}
\label{bcEM}
\mathbf{\hat z}\times {\bf E}'|_{\rm mirror}=0,\;\;\; \mathbf{\hat z}\cdot {\bf B}'|_{\rm mirror}=0.
\end{equation}
 
It is useful to decompose the fields into transverse electric (TE) and transverse magnetic (TM) polarizations (where `transverse' 
means perpendicular to the incidence plane defined by $ \mathbf{\hat z}$  and the propagation direction). 
For the TE component, we define the vector potential in the usual way: 
\(
{\bf E}^{(\rm TE)}=-\partial_t {\bf A}^{(\rm TE)},
\)
\(
{\bf B}^{(\rm TE)}=\boldsymbol{\nabla}\times {\bf A}^{(\rm TE)}
\)
under the Coulomb gauge $\boldsymbol{\nabla}\cdot  {\bf A}^{(\rm TE)}=0.$
Since ${\bf A}^{(\rm TE)}\cdot \mathbf{\hat z}=0,$  ${\bf A}^{(\rm TE)}$ is 
invariant under  the Lorentz boost from the co-moving frame to the laboratory frame. 
The resulting b.c. is then similar to  (\ref{Dirichlet}):
\begin{equation}
{\bf A}^{(\rm TE)}(x,y,q(t),t)={\bf 0}.
\label{bcAte}
\end{equation}

On the other hand, 
 $ {\bf A}^{(\rm TM)}$ has a component along the $z$-axis, so that 
the Coulomb gauge is no longer invariant under the Lorentz boost, resulting in complicated b.c.  also involving the scalar potential. 
It is then convenient to 
define a new vector potential $\boldsymbol{\cal A}^{(\rm TM)}$ as
\[
{\bf E}^{(\rm TM)}=\boldsymbol{\nabla}\times \boldsymbol{\cal A}^{(\rm TM)},\;\;\;\;{\bf B}^{(\rm TM)}=\partial_t \boldsymbol{\cal A}^{(\rm TM)}
\]
under the gauge $\boldsymbol{\nabla}\cdot \boldsymbol{\cal A}^{(\rm TM)}=0.$
Like ${\bf A}^{(\rm TE)}$, $\boldsymbol{\cal A}^{(\rm TM)}$ is also invariant under Lorentz boosts along the $z$-axis. 
From 
(\ref{bcEM}), one derives that $\boldsymbol{\cal A}^{(\rm TM)}$ satisfies a Neumann b.c.  at the co-moving frame, yielding 
\begin{equation}
\left[ \partial_z + \dot{q}(t)\partial_t+{\cal O}(\dot{q}^2)\right]\boldsymbol{\cal A}^{(\rm TM)}(x,y,q(t),t)={\bf 0} .
\label{bcAtm}
\end{equation}

The 
condition of perfect reflectivity then results in two independent problems: a Dirichlet b.c.  for TE modes, and 
a Neumann b.c. in the instantaneously co-moving frame for  TM modes.   The TM contribution turns out to be 11 times larger than the TE one, which 
coincides with (\ref{3scalar}). The resulting dissipative force is then \cite{Neto94}
\begin{equation}
\label{3EM}
f(t) = -\frac{\hbar A  q^{(5)}(t)}{30 \pi^2 c^4}. \hspace{30pt} (3\mbox{D, electromagnetic})
\end{equation}

As in the 1D case, the dissipative Casimir force plays the role of a radiation reaction force, associated with the emission of photon pairs with wave-vectors  
satisfying the conditions $|{\bf k}_1|+|{\bf k}_2|=\Omega/c$ and ${\bf k}_1{}_{\parallel}=-{\bf k}_2{}_{\parallel}$ from translational symmetry parallel to the 
plane of the mirror. 
The angular distribution of emitted photons displays an interesting correlation with polarization: TE photons are preferentially emitted near the normal direction,
whereas TM ones are preferentially emitted at larger angles, near a grazing direction if the frequency is smaller than $\Omega/2$ \cite{Neto96}.  
 
Results beyond the model of perfect reflection were obtained in Ref. \cite{Barton96A} for a dielectric half-space (see also Ref. \cite{Gutig98} for 
the angular and frequency spectra of emitted photons). In this case, there is also photon emission 
(and the associated dissipative radiation reaction force)
if the dielectric mirror moves sideways, or if a dielectric sphere rotates around a diameter \cite{Barton96B}. 
We will come back to this type of arrangement when discussing non-contact quantum friction 
in the next section.
 
To conclude this section, we compute the total photon production rate for a perfectly reflecting oscillating mirror directly from (\ref{3EM}). 
By energy conservation,
the total radiated energy is the negative of the work done on the mirror by the dissipative Casimir force:
\begin{equation}
\label{energy_conservation}
E = - \int_{-\infty}^{\infty} f(t) q^{(1)}(t) dt.
\end{equation}
We evaluate the integral in (\ref{energy_conservation})  using 
the result (\ref{3EM})
for an oscillatory motion of frequency $\Omega$ and amplitude $q_0$ exponentially damped over a time scale 
$T\gg 1/\Omega:$
\(
E = \hbar T  A q_0^2\Omega^6/(120\pi^2c^4).
\)
Since the spectrum is symmetrical with respect to the frequency $\Omega/2,$ we can derive the number of photons $N$ from the radiated energy
using the relation
 $E= N \hbar\Omega/2.$ 
The total photon production rate is then given by 
\begin{equation} 
\label{NT3D}
\frac{N}{T} =   \frac{1}{15}\,\frac{A}{\lambda_0^2}\, \left(\frac{v_{\rm max}}{c}\right)^2\,\Omega
\end{equation}
with 
$v_{\rm max}\equiv \Omega q_0$ and  
$\lambda_0\equiv 2\pi c/\Omega$ representing the typical scale of the relevant wavelengths. 
With  $v_{\rm max}/c\sim 10^{-7},$ $\Omega/(2\pi)\sim  10$ GHz and $A\sim \lambda_0^2\sim 10\,{\rm cm}^2,$ 
we find $N/T\sim 10^{-5}$ photons/sec or approximately one photon pair every two days! 

The dynamical Casimir effect is clearly 
very small for a single oscillating mirror.
Adding a second parallel plane mirror, the photon production rate is enhanced by a factor $(\Omega L/c)^{-2}\sim 10^6$  
for separation distances $L$ in the sub-micrometer range \cite{Mundarain98}. But at such short distances, finite conductivity of the metallic 
plates, not considered so far, 
is likely to 
reduce the photon production rate.  In this type of arrangement, a much larger effect is obtained by considering the shear motion of one plate relative to the other 
(instead of a relative motion along the normal direction). As discussed in the next section, 
because of finite conductivity a large friction force is predicted  at short distances, 
which results from the creation of pairs of excitations  inside the metallic medium \cite{Pendry1997, Pendry2010}. 
As for the emission of photon pairs, 
the orders of magnitude are  more promising when considering a closed  cavity with moving walls, to be discussed in Section 4. 

\section{Quantum Friction}

There is an intimate connection between the dynamical Casimir effect and the possibility that electrically neutral bodies in relative motion may experience
non-contact friction due to quantum vacuum fluctuations, the so-called ``quantum friction". As we have discussed in the previous section, dielectric
bodies in accelerated motion radiate Casimir photons. Shear motion of two bodies, even at constant relative speed, can also radiate energy.
Just as in the case of a single accelerated mirror in a vacuum, shear motion cannot be removed by a change of reference frame.
A frictional force between two perfectly smooth parallel planes shearing against each other with a relative velocity ${\bf v}$ results from the exchange of
photons between the two surfaces. These photons carry the information of the motion of one surface to the other one, and as a result linear momentum 
is exchanged between the plates, leading to friction. 

In order to illustrate the physics of quantum friction we will follow here an approach due to Pendry \cite{Pendry1997} who considered the simplest case of
zero temperature and the non-retarded (van der Waals) limit. The nice feature of this approach is that it manifestly connects to 
the intuitive picture of motion-induced (virtual) photons as mediators of momentum exchange between the shear surfaces. A dielectric surface,
although electrically neutral, experiences quantum charge fluctuations, and these have corresponding images on the opposing dielectric surface. Since the surfaces
are in relative parallel motion, the image lags behind the fluctuating charge distribution that created it, and this results in a frictional van der Waals force. Note that for
ideal perfect metals, the image charges arrange themselves instantaneously (do not lag behind), and therefore no quantum friction is expected in this case. 

We model each of the dielectric surfaces as a continuum of oscillators with Hamiltonian
\begin{equation}
\hat{H}_{\alpha} = \sum_{{\bf k} j} \hbar \omega_{\alpha; {\bf k} j} ( \hat{a}^{\dagger}_{\alpha; {\bf k} j} \hat{a}_{\alpha; {\bf k} j} + 1/2 ) ,
\end{equation}
where $\hat{a}^{\dagger}_{\alpha; {\bf k} j}$ and  $\hat{a}_{\alpha; {\bf k} j}$ are creation and annihilation bosonic operators associated with
the upper ($\alpha=u$) or lower ($\alpha=l$) plate. Each mode on each surface is defined by ${\bf k}$, which is a wave-vector parallel to the planar surface, and by $j$, which
denotes degrees of freedom perpendicular to the surface. Following Pendry we restrict ourselves to the non-retarded limit (very long wavelengths for the EM
modes). In this limit the EM field is mainly electrostatic, only the static TM polarization matters for a dielectric surface (since the static TE field
is essentially a magnetic field that does not interact with the non-magnetic surface), and the intensity decays exponentially from the surfaces (evanescent
fields).  The coupling between the oscillator modes belonging to different surfaces is mediated by the EM field, and it is assumed to be a position-position
interaction of the form
\begin{equation}
\hat{H}_{\rm int} (t) = \sum_{{\bf k} j j'}  C_{{\bf k} j j'} (d) \hat{x}_{u; {\bf k} j} \otimes  \hat{x}_{l; -{\bf k} j'} \; e^{-i k_x v t}.
\label{pertubation_quantumfriction}
\end{equation}
In the non-relativistic limit the effect of the surfaces shearing  with speed $v$ along the $x$ direction is contained in the last exponential factor.
This type of Hamiltonian follows from the
effective electrostatic interaction between the fluctuating charges in the dielectrics and its expansion to lowest order in the displacement of each oscillator
from its equilibrium position (equivalently, it also follows from the non-retarded and static limit of the dipole-dipole interactions between fluctuating dipoles
in each surface). The coupling factors $C_{{\bf k} j j'} (d)$ can be obtained by analyzing how each oscillator dissipates energy into the vacuum gap. This
is done in Ref. \cite{Pendry1997} in two ways, by invoking a scattering type of approach relating the fields at the interphases with reflection amplitudes,
and by considering how the fluctuating charge distributions in each surface dissipate energy. The result is 
$C_{{\bf k} j j'} (d)= (\beta_{{\bf k} j} \beta_{{\bf k} j'} / 2 k \epsilon_0) e^{- | {\bf k} | d}$, where
\begin{equation}
\beta^2_{{\bf k} j } = \left( \frac{dN}{d\omega} \right)^{-1} \; \frac{4 k \omega \epsilon_0}{\pi} \; {\rm Im} 
\left[ \frac{\epsilon(\omega) -1}{\epsilon(\omega)+1} \right] .
\label{betas}
\end{equation}
Note the exponential decay due to the evanescent nature of the EM field. In this equation $dN/d\omega$ is the density of oscillator modes at frequency $\omega$
and $\epsilon(\omega)$ is the complex dielectric permittivity of the plates (assumed to be identical). Although the interaction Hamiltonian does not depend
explicitly on the quantized EM field (because this derivation is semiclassical), one can infer the quantum processes of  creation and absorption that take
place by expanding the product $\hat{x}_{u; {\bf k} j}  \otimes \hat{x}_{l; -{\bf k} j'}$:
\begin{equation}
\hat{x}_{u; {\bf k} j}  \otimes \hat{x}_{l; -{\bf k} j'} = - \frac{1}{2} [ \hat{a}^{\dagger}_{u,{\bf k} j} - \hat{a}_{u, {\bf k} j} ] \otimes [ \hat{a}^{\dagger}_{l,-{\bf k} j'} - \hat{a}_{l, {-\bf k} j'} ] .
\label{friction_2excitations}
\end{equation}
Imagine the system of the two dielectrics is  initially in the ground state at zero temperature, $|\psi(t=0) \rangle = |\psi_g\rangle_u  \otimes |\psi_g\rangle_l$, where 
$|\psi_g\rangle_{\alpha} = \prod_{{\bf k} j} |\psi_{g, {\bf k} j} \rangle_{\alpha}$ is the product of the harmonic oscillators' ground states for surface 
$\alpha$. Eq.(\ref{friction_2excitations}) implies that two motion-induced virtual photons created from an EM vacuum produce one excitation in each surface, i.e.,  there is a non-zero probability of transition to states 
$|1; {\bf k} j \rangle_u \otimes | 1; -{\bf k} j'\rangle_l$.
The transition probability can be computed using time-dependent perturbation theory for the perturbation $H_{\rm int}(t)$. To first order, the transition probability from
the ground state into each of these two-excitation states is
\begin{equation}
P_{{\bf k} j j'}(t) = \frac{ \beta^2_{{\bf k}j}  \beta^2_{{\bf k}j'}}{4 k^2 \epsilon^2_0} \frac{e^{-2 d |{\bf k}|}}{4 \omega_{u,{\bf k}j}  \omega_{l, -{\bf k} j'}}
\frac{ 4 \sin^2[(\omega_{u,{\bf k}j}  + \omega_{l,-{\bf k}j'} - k_x v) t/2] }{ (\omega_{u,{\bf k}j}  + \omega_{l,-{\bf k}j'} - k_x v)^2}.
\label{transition_quantumfriction}
\end{equation}
In the limit of large times ($t \rightarrow \infty$) we use that $\sin^2(\Omega t/2)/(\Omega/2)^2 \approx \pi t \delta(\Omega)$ (here $\delta(\Omega)$ is Dirac's delta
function), and therefore the transition probability grows linearly in time. We can find the frictional force equating the frictional work $F_x v$ with the rate of change
in time of the energy of the excitations, namely
\begin{equation}
F_x v = \frac{dU}{dt} = \sum_{{\bf k} j j'} \hbar (\omega_{u, {\bf k} j} + \omega_{l, -{\bf k} j'} ) \frac{d P_{{\bf k} j j'}}{dt}  ,
\end{equation}
and the r.h.s. is time-independent since the transition probabilities grow linearly in time. 
Using the expression (\ref{betas}) for $\beta^2_{{\bf k}j}$ 
 and the transition probabilities (\ref{transition_quantumfriction}) at large times,
and
writing the sums over the dielectric degree of freedom $j$ as
$\sum_j = \int_0^{\infty} d\omega dN/d\omega$ (and similarly for $j'$),  one finally 
obtains the following expression for the frictional force 
\begin{equation}
F_x = \frac{\hbar}{\pi} \int \frac{d^2 {\bf k}}{(2 \pi)^2}  k_x e^{-2 |{\bf k}| d}  \int_0^{k_x v} d\omega \;  {\rm Im} \left[ \frac{\epsilon(\omega)-1}{\epsilon(\omega)+1} \right] \; 
{\rm Im} \left[ \frac{\epsilon(k_x v - \omega)-1}{\epsilon(k_x v - \omega)+1} \right] .
\end{equation}

In the literature there are  other more rigorous approaches to calculate quantum friction that go beyond the non-retarded quasi-static limit considered above, and that
can take into account effects of relativistic motion as well as finite temperature.
One of these approaches \cite{Volotikin1999} follows the spirit of the 
Lifshitz-Rytov theory \cite{Lifshitz1956}, considering the fluctuating electromagnetic (EM) field as a {\it classical} field whose stochastic fluctuations satisfy
the fluctuation-dissipation relation that relates the field fluctuations with the absorptive part of the dielectric response of the plates. The EM field is a solution
to Maxwell's equations with classical fluctuating current densities on the plates as source fields, and it satisfies the usual EM b.c. imposed
on the comoving reference frames on each plate. The relation between the EM fields in different frames is obtained via Lorentz transformations.
An alternative full quantum-mechanical approach considers the {\it quantum} EM field in interaction with (quantized) noise polarizations and noise currents within
the plates \cite{Buhmann2007}. As before, the fields in each reference frame are related by Lorentz transformations. In this approach the quantum expectation value of
the noise currents is given by the (quantum) fluctuation-dissipation relation.  

Quantum friction can also happen for neutral atoms moving close to surfaces. The theoretical methods to compute the frictional force in these cases are similar to the
surface-surface quantum friction, and we refer the reader to some of the relevant works \cite{Dedkov2002, Hu2004, Scheel2009}.

\section{Resonant photon creation in time dependent cavities}

As mentioned  in the Introduction, photon creation can be enhanced
in closed cavities: if the external time dependence involves a frequency that is 
twice the frequency of a mode of the electromagnetic field in the (unperturbed) cavity,
parametric amplification produces a large number of photons. As we will see, under 
certain circumstances (ideal three dimensional cavities with non equidistant frequencies 
in the spectrum) the number of photons in the resonant mode may grow exponentially.
Parametric amplification can take place by changing the length of the cavity with a 
moving surface, but also by changing its effective length through
time dependent electromagnetic properties of the cavity. 

In order to simplify the notation,
in this section we will use the natural units $\hbar =c=1$.

\subsection{Dynamical Casimir effect in 1D cavities}

As in Section 2, we start with a  massless real scalar field in a 1D cavity
with one mirror fixed at $x=0$ and the other performing an oscillatory
motion 
\begin{equation}
L(t)=L_0 [1+\epsilon \sin (\Omega t)]\,  , 
\label{trajmirror}
\end{equation}
where $\Omega$ is the external
frequency
and $\epsilon\ll 1$. As we will be mainly concerned with situations where
$\Omega L_0 = O(1)$, the maximum velocity of the mirror 
will be of order $\epsilon$, and therefore small values of $\epsilon$ 
correspond to a non-relativistic 
motion of the mirror.  We shall
assume that the oscillations begin at $t=0$, end at $t=T$,
and that $L(t)=L_0$ for $t<0$ and $t>T$.  The
scalar field $\phi(x,t)$ satisfies the wave equation $\Box\phi=0$
and Dirichlet b.c.  $\phi(x=0,t)=\phi(L(t),t)=0$.
When the mirror is at rest, the eigenfrequencies
are multiples of the fundamental frequency $\pi/L_0$. Therefore,
in order to analyze resonant situations we will assume that
$\Omega=q \pi / L_0$, $q=1,2,3,....$.

\begin{figure}[b]
\begin{center}
\includegraphics[scale=1]{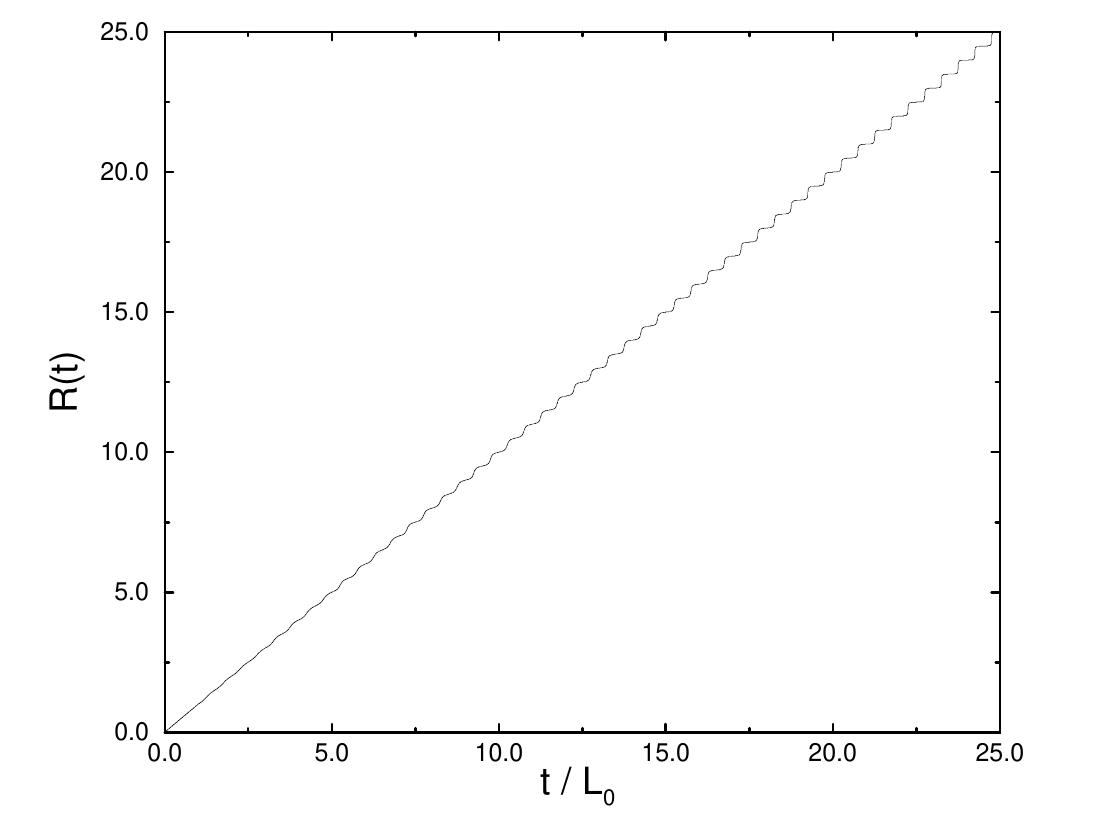}
\end{center}
\caption{$R(t)$ vs. $t/L_0$ as given by Eq.(\ref{finalrg}). 
The values of the parameters are $q=4$ and $\epsilon=0.01$.}
\label{stair}
\end{figure}

Inside the cavity we can write
\begin{equation}
\phi(x,t)= \sum_{k=1}^{\infty} \left[ a_k \psi_k(x,t) + a_k^{\dagger} \psi_k^{*}(x,t) \right] ,
\end{equation}
where the mode functions $\psi_k(x,t)$ are positive frequency modes
for $t<0$, and $a_k$ and $a_k^{\dagger}$ are time-independent
bosonic annihilation and creation operators, respectively. The field equation is automatically
verified by writing the modes in terms of Moore's function $R(t)$ \cite{Moore70} as
\begin{equation}
\psi_k(x,t)=\frac{i}{\sqrt{4 \pi k}} \left( e^{-i k \pi R(t+x)}  - e^{-i k \pi R(t-x)} \right) .
\label{modesD}
\end{equation}
The Dirichlet boundary condition is satisfied provided that $R(t)$ satisfies Moore's equation
\begin{equation}
R(t+L(t))-R(t-L(t))=2 .
\label{mooreq}
\end{equation}
These simple expressions for the modes of the field are due to conformal invariance, a symmetry for massless fields 
in one spatial dimension. 

\begin{figure}[b]
\begin{center}
\includegraphics[scale=1]{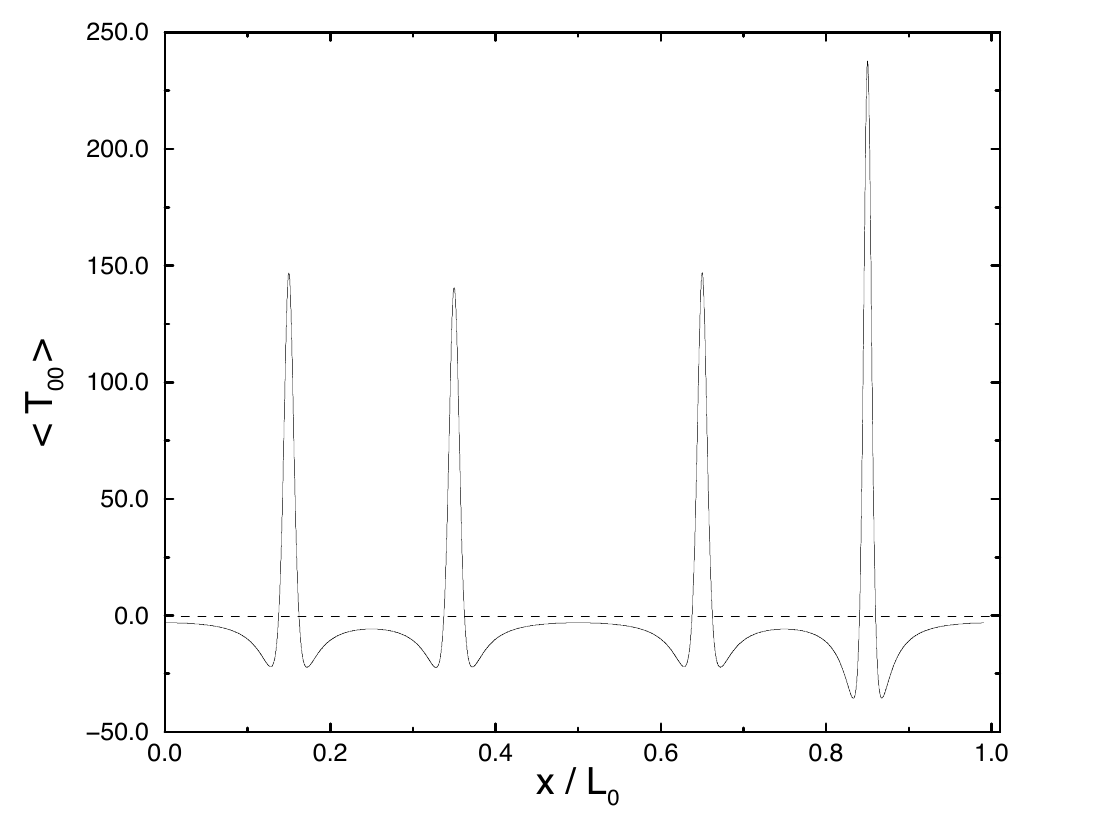}
\end{center}
\caption{ Energy density profile between plates for fixed time $t/L_0=20.4$ for
the $q=4$ case. The amplitude coefficient is $\epsilon=0.01$}
\label{density}
\end{figure}

The solution to the problem involves finding a solution
$R(t)$ in terms of the prescribed motion $L(t)$. For $t<0$ the positive frequency modes are given by $R(t)={t/L_0}$ for $- L_0 \leq t \leq L_0$, which is indeed a solution to Eq.(\ref{mooreq}) for $t<0$. For $t>0$, Eq.(\ref{mooreq}) can be solved, for example, 
using a perturbative expansion in $\epsilon$ similar to the one employed in Section 2 for a single mirror. However, as the external frequency is tuned with the unperturbed modes of
the cavity, in general there will be resonant effects, which produce secular terms proportional to 
$\epsilon^m(\Omega t)^n$ with $m\leq n$. Thus this expansion is valid only for short times $\epsilon \Omega t \ll 1$.
It is possible to obtain a non-perturbative solution of Eq.(\ref{mooreq})
using Renormalization Group (RG) techniques \cite{Dalvit98}. The RG-improved solution automatically adds the most
secular terms, $(\epsilon\Omega t)^n$, to all orders in $\epsilon$, and is valid for longer times $\epsilon^2 \Omega t \ll 1$. 
The RG-improved solution is \cite{Dalvit98,Dodonovetal93}
\begin{equation}
R(t) = \frac {t}{L_0} - \frac{2}{\pi q} {\mbox{Im}} \ln \left[
1 + \xi + (1-\xi) e^{\frac{i q \pi t}{L_0}} \right] ,
\label{finalrg}
\end{equation}
where $\xi = \exp[ (-1)^{q+1} \pi q \epsilon t / L_0]$. As shown in Fig.\ref{stair}, the function $R(t)$ develops a staircase shape for long times \cite{Dalvit98,Petrov2005}. Within
regions of $t$ between odd multiples of $L_0$ there appear $q$ jumps, located at values of $t$ satisfying $\cos(q \pi t / L_0)= \mp 1$, the upper sign corresponding to even values of $q$ and the lower one to odd values of $q$.

The vacuum expectation value of the energy density of the field is given by \cite{Fulling76}
$\langle T_{00}(x,t) \rangle = -f(t+x)-f(t-x)$,
where
\begin{equation}
f = \frac{1}{24 \pi} \left[ \frac{R'''}{R'} - \frac{3}{2} \left(
\frac{R''}{R'} \right)^2 + \frac{\pi^2}{2} (R')^2  \right] .
\label{EFE}
\end{equation}
For $q=1$ (``semi-resonant" case) no exponential amplification of
the energy density is obtained, whereas for  $q \geq 2$
(``resonant" cases)  the energy density grows exponentially in the
form of $q$ traveling wave packets which become narrower and
higher as time increases (see Fig.\ref{density}). Note that,  as the energy density involves the derivatives of 
the function $R(t)$,  there is one peak for each jump of $R(t)$.

The number of created particles can
be computed from the solution given by Eq.(\ref{finalrg}).  Photons are
created resonantly in all modes with $n=q+2j$, with $j$ a
non-negative integer. This is due to the fact that the spectrum of
a one dimensional cavity is {\it equidistant}: although the
external frequency resonates with a particular eigenmode of the
cavity, intermode coupling produces resonant creation in the other
modes. At long times,  the number of photons in each mode grows
linearly in time, while the total number of photons grows quadratically and the total energy inside 
the cavity grows
exponentially \cite{Dodonov96}.  These different behaviors are due
to the fact that the number of excited modes, i.e. the number of modes that
reach a growth linear in time, increases exponentially.  

The production of massless particles in one dimensional cavities has been
analyzed numerically in Ref. \cite{Ruser05}. As expected, the numerical evaluations are in 
perfect agreement with the analytic results described above in the case of small amplitudes
$\epsilon\leq 0.01$.

The force on the moving mirror can be computed as the discontinuity of  $\langle T_{11}\rangle$
at $x=L(t)$. The force produced
by the outside field is  one -half the expression derived in Section 2. This is much smaller than the intracavity contribution and will be neglected. Therefore
\begin{equation}
\langle F \rangle \approx \langle T_{11}(L(t),t) \rangle = \langle T_{00}(L(t),t) \rangle\, ,
\end{equation} 
where the energy momentum tensor is evaluated {\it inside} the cavity.
This expression reproduces the usual attractive result when the mirror is at rest ($t<0$)
\begin{equation}
\langle F \rangle=-\frac{\pi}{24L_0^2}\, ,
\end{equation}
that is, the static Casimir effect in 1D. 
However, at long times it becomes an exponentially increasing pressure due to 
the presence of real photons in the cavity \cite{Dodonov96}.

All this treatment can be extended to the case of Neumann b.c.
$n^{\mu}\partial_{\mu}\phi\vert_{\rm mirror}=0$, where $n^{\mu}$ is a
unit two-vector perpendicular to the trajectory of the mirror. The modes
of the field can be written in terms of Moore's function $R(t)$ as
\begin{equation}
\phi_k(x,t)=\frac{1}{\sqrt{4 \pi k}} \left( e^{-i k \pi R(t+x)}  +
e^{-i k \pi R(t-x)} \right) . \label{modesN}
\end{equation}
Note the change of sign between Eq.(\ref{modesD}) for Dirichlet
modes, and Eq.(\ref{modesN}) for Neumann modes. 
The spectrum of 
motion-induced photons is the same for
both Dirichlet and Neumann b.c. \cite{Alves03},
but not for the mixed configuration with one Dirichlet mirror and one Neumann mirror \cite{Alves06}.
The one dimensional DCE has also been investigated for cavities with Robin b.c. \cite{Farina10}.

\subsection{Photon creation in 3D cavities}

The one dimensional DCE with Dirichlet b.c.
described in the previous section 
is not only of academic interest: it describes photon creation for the 
TEM modes in a 3D cylindrical cavity with a non-simply connected section \cite{ Crocce05}.
However, in order to analyze TE and TM modes in general 3D cavities, 
a new approach is needed, since conformal invariance is no longer useful in 3D.

We shall first describe in some detail the simpler case of a scalar field in a rectangular
cavity satisfying Dirichlet b.c.
\cite{Crocce01}, and then comment on the extension to the case
of the electromagnetic field in cylindrical cavities with an arbitrary section.

\subsubsection{Scalar field}

We consider a rectangular cavity formed by perfectly reflecting walls 
with dimensions $L_{x},L_{y},$ and $L_{z}$. The wall placed at $z=L_{z}$ 
is at rest for $t<0$ and begins to move following a given trajectory, 
$L_{z}(t)$, at $t=0$. We assume this trajectory as prescribed for 
the problem (not a dynamical variable) and that it works as a time-dependent 
boundary condition for the field.
The field $\phi(\mathbf{x},\mathit{t})$ satisfies the wave equation 
$\Box\phi=0$, and the b.c. 
$\phi|_{\rm walls}=0$ for all times.
The Fourier expansion of the field for an arbitrary moment of time, 
in terms of creation and annihilation operators, can be written as

\begin{equation}
\phi(\mathbf{x},\mathit{t})=\sum_{\mathbf{n}}
\hat{a}_{\mathbf{n}}^{\scriptscriptstyle{\rm in}}
u_{\mathbf{n}}(\mathbf{x},\mathit{t}) + {\rm H.c.} ,
\label{field}
\end{equation} 
where the mode functions $u_{\mathbf{n}}(\mathbf{x},\mathit{t})$ form
a complete orthonormal  set of solutions of the wave equation 
with vanishing b.c.. 

When $t\leq0$ (static cavity) each field mode is 
determined by three positive integers $n_{x},n_{y}$ and $n_{z}$, namely

\begin{eqnarray}
u_{\mathbf{n}}(\mathbf{x},\mathit{t}<0)&=&{1\over\sqrt{2\omega_
{\mathbf{n}}}}\sqrt{\frac{2}{L_{x}}}\sin\left(\frac{n_{x}\pi}
{L_{x}} x\right)\sqrt{\frac{2}{L_{y}}}
\sin\left(\frac{n_{y}\pi}{L_{y}} y\right)
\nonumber\\
&\times&\sqrt{\frac{2}{L_{z}}}
\sin\left(\frac{n_{z}\pi}{L_{z}} z\right)
e^{-i\omega_{\mathbf{k}}t} ,
\label{expest}
\end{eqnarray}
with $\omega_{\mathbf{n}}=\pi\sqrt{\left(\frac{n_{x}}{L_{x}}\right)^{2} \!+ 
\left(\frac{n_{y}}{L_{y}}\right)^{2}\! + 
\left(\frac{n_{z}}{L_{z}}\right)^{2}}$ .

When $t>0$ the boundary condition on the moving wall becomes 
$\phi(x,y,z=L_{z}(t),t)=0$. In order to satisfy it we expand the mode 
functions in Eq.(\ref{field}) with respect to an \textit{instantaneous basis}  
\cite{Law94}

\begin{eqnarray}
u_{\mathbf{n}}(\mathbf{x},\mathit{t}>0)&=&\sum_{\mathbf{m}}
Q_{\mathbf{m}}^{(\mathbf{n})}(t)\sqrt{\frac{2}{L_{x}}}
\sin\left(\frac{m_{x}\pi}{L_{x}} x\right)
\sqrt{\frac{2}{L_{y}}}\sin\left(\frac{m_{y}\pi}{L_{y}} y\right)\nonumber\\
&\times&\sqrt{\frac{2}{L_{z}}}\sin\left(\frac{m_{z}\pi}{L_{z}(t)} z\right)
=\sum_{\mathbf{m}}Q_{\mathbf{m}}^{(\mathbf{n})}(t)\,
\varphi_{\mathbf{m}}(\mathbf{x},\mathit{L}_{z}(t)) ,
\label{exp}
\end{eqnarray}
with the initial conditions

\begin{equation}
Q_{\mathbf{m}}^{(\mathbf{n})}(0)={1\over\sqrt{2\omega_{\mathbf{n}}}}\,
\delta_{\mathbf{m},\mathbf{n}}  \ ,\ \ 
\dot{Q}_{\mathbf{m}}^{(\mathbf{n})}(0)=-i
\sqrt{\omega_{\mathbf{n}}\over2}\,\delta_{\mathbf{m},\mathbf{n}} .
\end{equation}
In this way we ensure that, as long as $L_{z}(t)$ and 
$\dot{L}_{z}(t)$ are continuous at $t=0$,
each field mode and its time derivative are also continuous functions. 
The expansion in Eq.(\ref{exp}) for the field modes must be a solution of the 
wave equation. Taking into account that the $\varphi_{\mathbf{k}}$'s 
form a complete and orthonormal set and that they depend on $t$ only 
through $L_{z}(t)$, we obtain a set of (exact) coupled 
equations for $Q_{\mathbf{m}}^{(\mathbf{n})}(t)$\cite{Crocce01}:
\begin{eqnarray}
\ddot{Q}_{\mathbf{m}}^{(\mathbf{n})}+\omega_{\mathbf{m}}^{2}(t)\,
Q_{\mathbf{m}}^{(\mathbf{n})}&=& 2\lambda(t)\sum_{\mathbf{j}}
g_{\mathbf{mj}}\,\dot{Q}_{\mathbf{j}}^{(\mathbf{n})}+\dot{\lambda}(t)
\sum_{\mathbf{j}}g_{\mathbf{mj}}\,Q_{\mathbf{j}}^{(\mathbf{n})}\nonumber\\
&+&\lambda^{2}(t)\sum_{\mathbf{j,l}}g_{\mathbf{lm}}\,g_{\mathbf{lj}}\,
Q_{\mathbf{j}}^{(\mathbf{n})} ,
\label{ecacop}
\end{eqnarray}
where 
\begin{equation}
\omega_{\mathbf{m}}(t)=\pi\sqrt{\left(\frac{m_{x}}{L_{x}}\right)^{2} \,+ 
\left(\frac{m_{y}}{L_{y}}\right)^{2}\, + \left(\frac{m_{z}}{L_{z}(t)}\right)^{2}}
\ \ ; \ \ \lambda(t)=\frac{\dot{L}_{z}(t)}{L_{z}(t)} .
\end{equation}
The coefficients $g_{\mathbf{mj}}$ are defined by
\begin{equation}
g_{\mathbf{mj}}=-g_{\mathbf{jm}} =L_{z}(t)\int_{0}^{L_{z}(t)}dz\ 
\frac{\partial\varphi_{\mathbf{m}}}{\partial L_{z}}\,\varphi_{\mathbf{j}}.
\end{equation}

The annihilation and creation operators $\hat{a}_{\mathbf{m}}^
{\scriptscriptstyle{\rm in}}$ and $\hat{a}^{\dag\,\scriptscriptstyle{\rm in}}_
{\mathbf{m}}$ correspond to the particle notion in the `in' 
region ($t<0$). If the wall stops for $t>t_{\rm final}$, we can define a new 
set  of operators, $\hat{a}_{\mathbf{m}}^{\scriptscriptstyle{\rm out}}$ and 
$\hat{a}^{\dag\,\scriptscriptstyle{\rm out}}_{\mathbf{m}}$, associated with
the particle notion in the `out' region ($t>t_{\rm final}$). 
These two sets of operators are connected by means of the Bogoliubov 
transformation
\begin{equation}
\hat{a}_{\mathbf{m}}^{\scriptscriptstyle{\rm out}}=
\sum_{\mathbf{n}} ( \hat{a}_{\mathbf{n}}^{\scriptscriptstyle{\rm in}}\,
\alpha_{\mathbf{nm}}+\hat{a}^{\dag\,
\scriptscriptstyle{\rm in}}_{\mathbf{n}}\,\beta_{\mathbf{nm}}^{\star} ) .
\label{bog1}
\end{equation}
The coefficients $\alpha_{\mathbf{nm}}$ and $\beta_{\mathbf{nm}}$ 
can be obtained as follows. When the wall returns to its initial position 
the right hand side in Eq.(\ref{ecacop}) vanishes and the solution is
\begin{equation}
Q_{\mathbf{m}}^{(\mathbf{n})}(t>t_{\rm final})=
A_{\mathbf{m}}^{(\mathbf{n})}e^{i\omega_{\mathbf{m}}t}+
B_{\mathbf{m}}^{(\mathbf{n})}e^{-i\omega_{\mathbf{m}}t},
\label{sol}
\end{equation}
with $A_{\mathbf{m}}^{(\mathbf{n})}$ and $B_{\mathbf{m}}^{(\mathbf{n})}$ 
being some constant coefficients to be determined by the continuity 
conditions at $t=t_{\rm final}$. 
Inserting Eq.(\ref{sol}) into Eqs.(\ref{field}) and (\ref{exp}) we obtain 
an expansion of $\phi$ in terms of 
$\hat{a}_{\mathbf{m}}^{\scriptscriptstyle{\rm in}}$ and 
$\hat{a}_{\mathbf{m}}^{\dag\,\scriptscriptstyle{\rm in}}$ for 
$t>t_{\rm final}$. Comparing this with the equivalent expansion in terms of 
$\hat{a}_{\mathbf{m}}^{\scriptscriptstyle{\rm out}}$ and 
$\hat{a}_{\mathbf{m}}
^{\dag\,\scriptscriptstyle{\rm out}}$ it is easy to see that
\begin{equation}
\alpha_{\mathbf{nm}}=\sqrt{2\omega_{\mathbf{m}}} 
B_{\mathbf{m}}^{(\mathbf{n})}\ \ , \ \ \beta_{\mathbf{nm}}=
\sqrt{2\omega_{\mathbf{m}}}\,A_{\mathbf{m}}^{(\mathbf{n})} .
\label{bog2}
\end{equation}

The amount of photons created in the mode $\mathbf{m}$ is the average value 
of the number operator 
$\hat{a}_{\mathbf{m}}^{\dag\,\scriptscriptstyle{\rm out}}
\hat{a}_{\mathbf{m}}^{\scriptscriptstyle{\rm out}}$ with respect to the 
initial vacuum state (defined through 
$\hat{a}_{\mathbf{m}}^{\scriptscriptstyle{\rm in}}
|0_{\scriptscriptstyle{\rm in}}\rangle=0$). With the help of 
Eq.(\ref{bog1}) and Eq.(\ref{bog2}) we find
\begin{equation}
\langle {\mathcal{N}}_{\mathbf{m}} \rangle=\langle 
0_{\scriptscriptstyle{\rm in}}\mid 
\hat{a}_{\mathbf{m}}^{\dag\,\scriptscriptstyle{\rm out}} 
\hat{a}_{\mathbf{m}}^{\scriptscriptstyle{\rm out}}
\mid 0_{\scriptscriptstyle{\rm in}} \rangle = 
\sum_{\mathbf{n}}2\omega_{\mathbf{m}}|A_{\mathbf{m}}^{(\mathbf{n})}|^{2}
\label{numerodefotones} .
\end{equation}
 
In the approach described so far we worked at the level of the dynamical equation
for the quantum scalar field. Alternatively, one can analyze the problem using the
{\it effective Hamiltonian method} developed in Ref. \cite{Schutzhold98}. 
The idea is the following. Assume that
a massless scalar field is confined within a time dependent volume and satisfies
Dirichlet b.c.. At the classical level, the field
can be expanded in terms of a basis of functions $f_{\alpha}({\bf x},t)$ that fulfill
the b.c. at each time, that is
\begin{equation}
\phi({\bf x},t)=\sum_{\alpha}q_{\alpha}(t)f_{\alpha}({\bf x},t)\, .
\end{equation}
For the rectangular cavities considered in this section these functions can be chosen
to be $\varphi_{\mathbf{m}}(\mathbf{x},\mathit{L}_{z}(t))$. Inserting this expansion 
into the Klein-Gordon Lagrangian, one ends up with a Lagrangian for the
generalized coordinates $q_\alpha(t)$, which is a quadratic function of 
$q_\alpha(t)$ and $\dot q_\alpha(t)$, i.e. it describes a set of coupled harmonic oscillators
with time dependent frequencies and couplings. This system can be quantized following the 
usual procedure, and the final results for the number of created 
photons are equivalent to those obtained in Eq. (\ref{numerodefotones}).


\subsubsection{Parametric amplification in 3D}

As in Section 4.1, we are interested in resonant situations where the number of 
photons created inside the cavity could be enhanced
for some specific external 
frequencies. So we study the trajectory given in Eq.(\ref{trajmirror}).
To first order in $\epsilon$, the equations for 
the modes Eq.(\ref{ecacop}) take the form
\begin{eqnarray}
\ddot{Q}^{({\bf n})}_{\bf m} + \omega^2_{\bf m} Q^{({\bf n})}_{\bf m} & =&
2 \epsilon \left(\frac{\pi m_z}{L_z}\right)^2 \sin(\Omega t) 
Q^{({\bf n})}_{\bf m}
- \epsilon \Omega^2 \sin(\Omega t) \sum_{\bf j} g_{{\bf m}{\bf j}} 
Q^{({\bf n})}_{\bf j} \nonumber\\
&+& 2 \epsilon \Omega \cos(\Omega t) \sum_{\bf j} g_{{\bf m}{\bf j}}
\dot{Q}^{({\bf n})}_{\bf j} +  O(\epsilon^2) .
\label{eqqk}
\end{eqnarray}

It is known that a naive perturbative solution of these equations in powers
of the displacement $\epsilon$ breaks down after a short amount of time, 
of order $(\epsilon\Omega)^{-1}$.  As in the 1D case discussed in the previous section,
this happens for those particular values of the external
frequency $\Omega$ such that there is
a resonant coupling with the eigenfrequencies of the static 
cavity. In this situation, to find a solution valid for 
longer times (of order $\epsilon^{-2}\Omega^{-1}$)
we proceed as follows. We  assume that
the solution of Eq.(\ref{eqqk}) is of the form
\begin{equation}
Q^{({\bf n})}_{\bf m}(t) = A^{({\bf n})}_{\bf m}(t) e^{i \omega_{\bf m} t}
+ B^{({\bf n})}_{\bf m}(t) e^{-i \omega_{\bf m} t} ,
\label{ansqk2}
\end{equation}
where the functions  $A^{({\bf n})}_{\bf m}$ and 
$B^{({\bf n})}_{\bf m}$ are slowly varying. In order to obtain
differential equations for them, we insert this ansatz
into Eq.(\ref{eqqk}) and  neglect second-order derivatives of 
$A^{({\bf n})}_{\bf m}$ and $B^{({\bf n})}_{\bf m}$. After multiplying the 
equation by $ e^{\pm i \omega_{\bf m} t}$
we average over the fast oscillations. The resulting equations are
\begin{eqnarray}
&&\frac{1}{\epsilon}{d A^{({\bf n})}_{\bf m}\over dt} =
-\frac{\pi^2 m_z^2}{2 \omega_{\bf m} L_z^2} B^{({\bf n})}_{\bf m}
\delta(2 \omega_{\bf m} -\Omega) \nonumber\\
&& +
\sum_{\bf j} (-\omega_{\bf j} + \frac{\Omega}{2})
\delta(-\omega_{\bf m} - \omega_{\bf j} + \Omega) 
\frac{\Omega}{2 \omega_{\bf m}} g_{{\bf k}{\bf j}} 
B^{({\bf n})}_{\bf j} \nonumber \\
&& + \sum_{\bf j} \left[
(\omega_{\bf j} + \frac{\Omega}{2}) 
\delta(\omega_{\bf m} - \omega_{\bf j} - \Omega) +
(\omega_{\bf j} - \frac{\Omega}{2}) 
\delta(\omega_{\bf m} - \omega_{\bf j} + \Omega)
\right]\nonumber\\
&&\times
\frac{\Omega}{2 \omega_{\bf m}} g_{{\bf m}{\bf j}} A^{({\bf n})}_{\bf j} ,
\label{ec1}
\end{eqnarray}
and 
\begin{eqnarray}
&&\frac{1}{\epsilon} {d B^{({\bf n})}_{\bf m}\over dt} =
-\frac{\pi^2 m_z^2}{2 \omega_{\bf m} L_z^2} A^{({\bf n})}_{\bf m}
\delta(2 \omega_{\bf m} -\Omega) \nonumber\\ &&+
\sum_{\bf j} (-\omega_{\bf j} + \frac{\Omega}{2})
\delta(-\omega_{\bf m} - \omega_{\bf j} + \Omega) 
\frac{\Omega}{2 \omega_{\bf m}} g_{{\bf m}{\bf j}} 
A^{({\bf n})}_{\bf j} \nonumber \\
&& + \sum_{\bf j} 
\left[
(\omega_{\bf j} + \frac{\Omega}{2}) 
\delta(\omega_{\bf m} - \omega_{\bf j} - \Omega) +
(\omega_{\bf j} - \frac{\Omega}{2}) 
\delta(\omega_{\bf m} - \omega_{\bf j} + \Omega)
\right]\nonumber\\ &&\times
\frac{\Omega}{2 \omega_{\bf m}} g_{{\bf m}{\bf j}} B^{({\bf n})}_{\bf j} ,
\label{ec2}
\end{eqnarray} 
where we used the notation $\delta(\omega)$ for the Kronecker $\delta$-function
$\delta_{\omega 0}$.

The method used to derive these equations is equivalent to the ``multiple scale analysis" \cite{MSA} and
to the slowly varying envelope approximation \cite{envelope}.
The  equations are non-trivial (i.e., lead to resonant 
behavior) if $\Omega=2 \omega_{\bf m} \label{cond0}$ (resonant condition). 
Moreover, there is intermode coupling between modes ${\bf j}$ and 
${\bf m}$ if any of the conditions $|\omega_{\mathbf{m}}\pm\omega_{\mathbf{j}}|=\Omega$ is satisfied.

We derived the equations for three dimensional cavities. It is easy to
obtain the corresponding ones for one dimensional cavities. The
conditions for resonance and intermode coupling are the same.
The main
difference is that for one dimensional cavities the
spectrum is equidistant. Therefore an infinite set of modes may be
coupled. 
For example, when the external frequency is $\Omega
=2 \omega_1$, the mode $m$ is coupled with the modes $m\pm 2$.
This has been extensively studied in the literature 
\cite{Dalvit98,Dodonov96,Soh,Dodonov96B}. 

In what follows we will be concerned with cavities with
non-equidistant spectrum.
Eqs.(\ref{ec1}) 
and (\ref{ec2}) present different kinds of 
solutions depending both on the mirror's frequency and the spectrum of the 
static cavity. In the simplest  `parametric resonance case'  the frequency of 
the mirror is twice the frequency of some unperturbed mode, say 
$\Omega=2\omega_{\mathbf{m}}$. In order to find $A_{\mathbf{m}}^{(\mathbf{n})}$ and 
$B_{\mathbf{m}}^{(\mathbf{n})}$ from Eq.(\ref{ec1}) and Eq.(\ref{ec2}) we 
have to analyze whether the coupling conditions 
$|\omega_{\mathbf{m}}\pm\omega_{\mathbf{j}}|=\Omega$ can be satisfied or not. 
If we set $\Omega=2\omega_{\mathbf{m}}$, the resonant mode $\mathbf{m}$ will 
be coupled to some other mode $\mathbf{j}$ only if 
$\omega_{\mathbf{j}}-\omega_{\mathbf{m}}=\Omega=2\omega_{\mathbf{m}}$. 
Clearly, the latter relation will be satisfied depending on the spectrum of 
the particular cavity under consideration.

Let us assume that this condition is not fullfilled. In this case, 
the equations for $A_{\mathbf{m}}^{(\mathbf{n})}$ and 
$B_{\mathbf{m}}^{(\mathbf{n})}$ can be easily solved and give
\begin{equation}
\langle {\mathcal{N}}_{\mathbf{m}} \rangle =   \sinh^{2}
\left[ \frac{1}{\Omega} \left( \frac{m_z\pi}{L_z} \right)^2\epsilon t_f
\right] .
\label{casododonov}
\end{equation}
In this uncoupled resonance case the average number of created 
photons in the mode ${\mathbf{m}}$ increases exponentially in time. Another way of looking at this particular 
situation
is to note that, neglecting the intermode couplings, the amplitude of the 
resonant mode satisfies the equation of an harmonic oscillator with time 
dependent frequency. For the particular trajectory given in Eq.(\ref{trajmirror}),
the dynamics of the mode is governed by Eq. (\ref{eqqk}) with $g_{{\bf m}{\bf j}}=0$ , that is
\begin{equation}
\ddot{Q}^{({\bf n})}_{\bf m} + \left [ \omega^2_{\bf m} -
2 \epsilon \left(\frac{\pi m_z}{L_z}\right)^2 \sin(\Omega t) 
\right ] Q^{({\bf n})}_{\bf m}=0\, ,
\label{eqMathieu}
\end{equation}
which is the well known Mathieu equation \cite{MSA}. The solutions to this equation
have an exponentially growing amplitude when $\Omega=2 \omega_{\bf m}$. 

There are simple situations in which there is intermode coupling.
For instance for a cubic cavity of size $L$, the fundamental mode $(1,1,1)$ is coupled
to the mode $(5,1,1)$ when the external frequency is  $\Omega=2 \omega_{111}$.
In this case the number of photons in each mode grows with a lower rate 
than that of the uncoupled case \cite{Crocce01}
\begin{equation}
\langle {\mathcal{N}}_{111} \rangle\simeq\langle {\mathcal{N}}_{511} \rangle\simeq 
e^{0.9\epsilon t_f/L}\,.
\end{equation}
We will describe some additional examples of intermode coupling in the next subsection, in the context of a
full electromagnetic model.

It is worth stressing that the creation of scalar particles in 3D cavities has been studied
numerically in Ref.\cite{Ruser06}. At long times,  the numerical results coincide with
the analytical predictions derived from Eqs.(\ref{ec1}) 
and (\ref{ec2}), both in the presence and absence 
of intermode coupling.

\subsubsection{The electromagnetic case}

The previous results have been generalized to the case of the electromagnetic field inside 
a cylindrical cavity, with an arbitrary transversal section \cite{ Crocce05}. 
Let us assume that the axis of the cavity is along 
the $z$-direction,  and that the caps are located at $z=0$ and $z=L_z(t)$. All 
the surfaces are perfect conductors.

When studying the electromagnetic field inside these cavities it is convenient
to express the physical degrees of freedom in terms of  the vector potentials  ${\bf  A}^{(\rm TE)}$ and 
$\boldsymbol{\cal A}^{(\rm TM)}$ 
introduced in Section 2.  These vectors can be written in terms of the so called
``scalar Hertz potentials" 
as  ${\bf  A}^{(\rm TE)} = {\bf\hat z}\times \nabla \phi^{\rm TE} $ and
$\boldsymbol{\cal A}^{(\rm TM)} = {\bf\hat z} \times \nabla\phi^{\rm TM}$.  For perfect reflectors the b.c. do not mix
TE and TM polarizations, and therefore  the electromagnetic field inside the cavity  can be described in terms of these
two {\it independent} scalar Hertz potentials: no crossed terms appear in Maxwell's Lagrangian or
Hamiltonian. 

The scalar Hertz potentials satisfy the Klein-Gordon equation. 
The b.c. of both potentials on the static walls of the cavity are
\begin{eqnarray}
\phi^{\rm TE} |_{z=0}=0   &;&
\frac{\partial \phi^{\rm TE}}{\partial n}\left|_{{\rm trans}}\right.=0  ,\\
\frac{\partial \phi^{\rm TM}}{\partial z}\left|_{z=0}\right.=0  &;&  \phi^{\rm TM} |_{{\rm trans}}=0 ,
\label{bcboth}
\end{eqnarray}
where $\partial /\partial n$ denotes  the normal derivative on the transverse boundaries. 
On the other hand, the b.c. on the moving mirror has been already discussed 
in Section 2 (see Eqs (\ref{bcAte}) and (\ref{bcAtm})). In terms of the Hertz potentials they read as
\begin{equation}
\left. \phi^{\rm TE} \right|_{z=L_z(t)} = 0 \;\; ; \;\;
\left. (\partial_z + \dot{L_z} \; \partial_t ) \phi^{\rm TM} \right|_{L_z(t)} = 0 .
\end{equation}
The energy of the electromagnetic field 
\begin{equation}
H=\frac{1}{8\pi}\int d^3x({\bf E}^2+{\bf B}^2)=H^{\rm TE}+H^{\rm TM} 
\end{equation}
can be written in terms of the scalar potentials as
\begin{equation}
H^{\rm (P)}=\frac{1}{8\pi}\int d^3x\left[\dot\phi^{\rm (P)}(-\nabla_\perp^2)\dot\phi^{\rm (P)}+
\phi^{\rm (P)'}(-\nabla_\perp^2)\phi^{\rm (P)'}+
\nabla_\perp^2\phi^{\rm (P)}\nabla_\perp^2\phi^{\rm (P)}\right]\,\, ,
\label{scalar energy}
\end{equation}
where dots and primes denote derivatives with respect to time and $z$ respectively.
The supraindex $\rm P$ corresponds to TE and TM and ${\bf \nabla}_\perp$ denotes the
gradient on the $xy$ plane.

The quantization procedure has been described in detail in previous papers
\cite{Crocce02,Crocce05,Hacyan90}. At any given time both
scalar Hertz potentials  can be expanded in terms of an instantaneous basis
\begin{equation}
\phi^{\rm (P)}({\bf x},t ) = \sum_{\bf n} a_{\bf n}^{\rm IN}
C_{{\bf n}}^{{\rm (P)}} u_{{\bf n}}^{{\rm (P)}}({\bf x}, t) +
{\rm c.c.} ,
\end{equation}
where $a_{\bf n}^{\rm IN}$ are bosonic operators that annihilate the IN vacuum state for $t<0$, and $C_{{\bf n}}^{{\rm (P)}}$ are
normalization constants that must be appropriately included to obtain the usual form of the electromagnetic Hamiltonian 
(\ref{scalar energy}) in terms of
annihilation and creation operators.  
 
For TE modes, the mode functions are similar to those of the scalar field satisfying Dirichlet b.c.
described in the previous section 
\begin{equation}
u_{{\bf n}}^{{\rm TE}} = \sum_{\bf m } Q^{({\bf n})}_{{\bf m},{\rm TE}}(t)  \sqrt{2/L_z(t)}
\sin\left( \frac{m_z \pi z}{L_z(t)} \right) v_{{\bf m}_{\perp}}({\bf x}_{\perp}).
\end{equation}
For TM modes. the choice of the instantaneous basis is less trivial and has been derived in detail
in Ref. \cite{Crocce02}
\begin{equation}
u_{{\bf n}}^{{\rm TM}} =
\sum_{\bf m} [ Q^{({\bf n})}_{{\bf m},{\rm TM}}(t)  + \dot{Q}^{({\bf n})}_{{\bf m},{\rm TM}}(t) g(z,t) ]
\sqrt{2/L_z(t)}
\cos\left( \frac{m_z \pi z}{L_z(t)} \right) r_{{\bf m}_{\perp}}({\bf x}_{\perp}) .
\end{equation}
Here the index ${\bf m} \neq 0$ is a vector of non-negative
integers.  The
function $g(z,t)=\dot{L}_z(t) L_z(t) \xi(z/L_z(t))$ (where
$\xi(z)$ is a solution to the conditions $\xi(0)=\xi(1)=\partial_z
\xi(0)=0$, and $\partial_z \xi(1)=-1$) appears when expanding the
TM modes in an instantaneous basis and taking the small $\epsilon$
limit. There are many solutions for $\xi(z)$, but all of them can
be shown to lead to the same physical results \cite{Crocce02}.
The mode functions $v_{{\bf m}_{\perp}}({\bf x}_{\perp})$ and
$r_{{\bf m}_{\perp}}({\bf x}_{\perp})$, are described below for
different types of cavities.

The mode functions $Q^{({\bf n})}_{{\bf m},{\rm TE/TM}}$ satisfy second order,
mode-coupled linear differential equations similar to Eq. (\ref{eqqk})
\cite{Crocce02}. As before,  for the ``parametric resonant case''
($\Omega = 2 \omega_{\bf n}$ for some ${\bf n}$) there is
parametric amplification.
Moreover,  for some particular geometries and sizes of the
cavities, different modes ${\bf n}$ and ${\bf m}$ can be coupled,
provided either of the resonant coupling conditions $\Omega = |
\omega_{\bf n} \pm \omega_{\bf m} |$ are met. When intermode
coupling occurs it affects the rate of photon creation, typically
resulting in a reduction of that rate.

The number of motion-induced photons with a given wavevector ${\bf n}$ and polarization 
TE or TM can be
calculated in terms of the Bogoliubov coefficients.
When the resonant coupling conditions are not met, the
different modes will not be coupled during the dynamics. As in the scalar case,  
the system can be described by a Mathieu equation (\ref{eqMathieu}) for a single mode. As a 
consequence, the number of motion-induced photons in that given mode will grow exponentially. The growth rate is different for TE and TM modes \cite{Crocce02}
\begin{equation}
\langle {\mathcal N}_{{\bf n},{\rm TE}}(t) \rangle = \sinh^2(\lambda_{{\bf n},{\rm TE}} \epsilon t)  \;\; ; \;\;
\langle {\mathcal N}_{{\bf n},{\rm TM}}(t) \rangle = \sinh^2(\lambda_{{\bf n},{\rm TM}} \epsilon t) ,
\end{equation}
where $\lambda_{{\bf n},{\rm TE}} = n_z^2 / 2 \omega_{\bf n}$ and $\lambda_{{\bf n},{\rm TM}} = (2 \omega^2_{\bf n} - n_z^2)/ 2
\omega_{\bf n}$. When both polarizations are present, the rate of growth for TM photons is larger than for TE photons, {\it i.e.}, $\lambda_{{\bf n},{\rm TM}} > \lambda_{{\bf n},{\rm TE}}$. As in the case of the scalar field, these equations are valid for 
$\epsilon^2\Omega t\ll 1$.

We describe  some specific examples:

{\it Rectangular section.} For a waveguide of length $L_z(t)$ and
transversal rectangular shape (lengths $L_x,L_y$), the TE mode
function is
\begin{equation}
v_{n_x,n_y}({\bf x}_{\perp}) = \frac{2}{\sqrt{L_x L_y}}  \cos \left( \frac{n_x \pi x}{L_x} \right)
\cos \left( \frac{n_y \pi y}{L_y} \right),
\end{equation}
with $n_x$ and $n_y$ non-negative integers that cannot be simultaneously zero.  
The spectrum is
\begin{equation}
\omega_{n_x,n_y,n_z} = \sqrt{(n_x \pi/L_x)^2 + (n_y \pi/L_y)^2 + (n_z \pi/L_z)^2} ,
\label{omegarect}
\end{equation}
with $n_z\geq 1$.
The TM mode function is
\begin{equation}
r_{m_x,m_y}({\bf x}_{\perp}) = \frac{2}{\sqrt{L_x L_y}} \sin \left( \frac{m_x \pi x}{L_x} \right)
\sin \left( \frac{m_y \pi y}{L_y} \right),
\end{equation}
where $m_x, m_y$ are positive  integers.   The spectrum is given by $\omega_{m_x,m_y,m_z}$, with $m_z\geq 0$.
 
Let us analyze the particular case of a cubic cavity of size $L$ under the parametric resonant condition
$\Omega= 2 \omega_{\bf k}$. The fundamental TE mode is doubly degenerate ($(1,0,1)$ and $(0,1,1)$) and uncoupled to other modes. 
The number of photons in these TE modes grows as $\exp (\pi\epsilon t\sqrt 2L)$.
The fundamental TM mode $(1,1,0)$ has the same energy  as the fundamental TE mode, and it is coupled to the TM mode $(1,1,4)$. Motion-induced TM photons are produced exponentially as $\exp(4.4 \epsilon t / L)$, much faster than TE photons.

{\it Circular section.} For a waveguide with a
transversal circular shape of radius $R$, the TE mode function is
\begin{equation}
v_{nm}({\bf x}_{\perp})=\frac{1}{\sqrt{\pi}} \frac{1}{R J_n(y_{nm}) \sqrt{1-n^2/y_{nm}^2}}
J_n \left( y_{nm}\frac{\rho}{R} \right)  e^{i n \phi} ,
\end{equation}
where $J_n$ denotes the Bessel function of {\it n}th order,
and $y_{nm}$ is the {\it m}th positive root of the equation
$J'_n(y)=0$. The eigenfrequencies are given by
\begin{equation}
\omega_{n,m,n_z}=\sqrt{\left(\frac{y_{nm}}{R}\right)^2+\left(\frac{n_z \pi}{L_z}\right)^2} ,
\label{spectrumcyl}
\end{equation}
where $n_z \ge 1$. The TM mode function is
\begin{equation}
r_{nm}({\bf x}_{\perp})=\frac{1}{\sqrt{\pi}} \frac{1}{R J_{n+1}(x_{nm})}
J_n \left( x_{nm} \frac{\rho}{R} \right) \ \ e^{i n \phi} ,
\end{equation}
where $x_{nm}$ is the {\it m}th root of the equation $J_n(x)=0$. The
spectrum is given by Eq.(\ref{spectrumcyl}) with $y_{nm}$ replaced by $x_{nm}$ and
$n_z \ge 0$. Denoting the modes by $(n,m,n_z)$, the lowest TE mode
is $(1,1,1)$ and has a frequency $\omega_{111}=(1.841/R)\sqrt{1+
2.912 (R/L_z)^2}$.  This mode is uncoupled to any other modes, and
the number of photons in this mode grows exponentially in time as
$\exp{(\pi \epsilon t / \sqrt{1+ 0.343 (L_z/R)^2} L_z)}$ when
parametrically excited.  The lowest TM mode $(0,1,0)$ is also
uncoupled and has a frequency $\omega_{010}=2.405/R$. The parametric
growth is $\exp{(4.81 \epsilon t/R)}$.
 For $L_z$ large enough ($L_z > 2.03 R$), the resonance frequency $\omega_{111}$ of the lowest
TE mode is smaller than that for the lowest TM mode. Then the
$(1,1,1)$ TE mode is the fundamental oscillation of the cavity.

\subsection{Time dependent electromagnetic properties}

From  a theoretical point of view,  it is possible to create photons from the vacuum not
only for a cavity with a moving mirror, but also when the electromagnetic properties of
the walls and/or the media inside the cavity change with time. Given the difficulties 
in a possible experimental verification of the DCE for moving mirrors,  the consideration of time dependent properties is not only of academic interest, but it is also relevant for the
analysis of the experimental proposals discussed in Section 5.

A setup that has attracted both theoretical and experimental attention is the possibility of using short laser pulses in order to produce periodic variations of the conductivity of a semiconductor layer placed inside a microwave cavity. The fast changes in the conductivity induce a periodic variation in the effective length of the cavity, and 
therefore the creation of photon pairs \cite{Yablonovitch89,Lozovik95}. This setup has been analyzed at the theoretical level \cite{Crocce04,Dodonov06, schutz,jap}, 
and there is an ongoing experiment aimed at the detection of the motion induced
radiation \cite{Braggio05} (see Section 5).

For the sake of clarity we discuss in detail the model of a massless scalar field
within a rectangular cavity with perfect conducting walls with
dimensions $L_x$, $L_y$, and $L_z$ described in Ref.\cite{Crocce04}. At the midpoint of the cavity
($x=L_x/2$)  there is a plasma sheet. We
model the conductivity properties  of such material
by a delta-potential with a time dependent strength $V(t)$. This is a time dependent
generalization of the model introduced in Ref.\cite{Barton95}. The strength of
the potential is given by
\begin{equation}
V(t)=4\pi\frac{e^2n(t)}{m^*}\,\, ,
\end{equation}
where $e$ is the electron charge, $m^*$ the electron's effective mass in the conduction band
and $n(t)$ the surface density of carriers. We assume that the irradiation of the 
plasma sheet produces changes in this quantity. 
The ideal limit of perfect conductivity corresponds to $V
\rightarrow \infty$, and $V \rightarrow 0$ to a `transparent'
material. The strength of the potential varies
between a minimum value, $V_0$, and a maximum $V_{\rm max}$. The
Lagrangian of the scalar field within the cavity is given by 
\begin{equation}
{\mathcal L}=\frac{1}{2} \partial_{\mu}\phi\partial^{\mu}\phi -
\frac{V(t)}{2} \delta(x-L_x/2) \phi^2, \label{themodel} 
\end{equation}
where
$\delta(x)$ is the one-dimensional Dirac delta function.
The use of an infinitely thin film is justified as long as the width of the slab is much 
smaller than 
the wavelengths of the relevant electromagnetic modes in the cavity. 
The
corresponding Lagrange equation reads, 
\begin{equation} (\nabla^2 -
\partial_t^2) \phi = V(t) \delta(x-L_x/2)\phi.
\label{fieldequation} 
\end{equation}

We divide the cavity into two regions: region I ($0\leq x \leq L_x/2$)
and region II ($L_x/2 \leq x \leq L_x$). Perfect conductivity at the edges of the cavity
imposes Dirichlet b.c. for the field.
The presence of the plasma sheet  introduces a discontinuity in the $x$-spatial derivative, while the field itself remains continuous,
\begin{eqnarray}
\phi_{\rm I}(x=L_x/2,t)&=&\phi_{\rm II}(x=L_x/2,t), \nonumber \\
\partial_x\phi_{\rm I}(x=L_x/2,t)-\partial_x \phi_{\rm II}(x=L_x/2,t)&=&-V(t) \phi(x=L_x/2,t).
\label{BC2}
\end{eqnarray}
We will consider a  set of solutions that satisfies
automatically all b.c.. 
\begin{equation}
\psi_{\bf m}({\bf x},t) = \sqrt{\frac{2}{L_x}}
\sin\left(k_{m_x}(t)\,x\right)\sqrt{\frac{2}{L_y}}
\sin\left(\frac{\pi m_y  y}{L_y}\right)\sqrt{\frac{2}{L_z}}
\sin\left(\frac{\pi m_z  z}{L_z}\right), \label{sol2}
\end{equation}
where $m_y, m_z$ are positive integers.
The function $\psi_{\bf m}$ depends on $t$ through  $k_{m_x}(t)$,
which is the $m_x$-th positive solution to the following
transcendental equation 
\begin{equation} 2 k_{m_x}
\tan^{-1}\left(\frac{k_{m_x}L_x}{2}\right)=- V(t) .
\label{trascendentalequation} 
\end{equation} 
To simplify the notation, in
what follows we will write $k_m$ instead of $k_{m_x}$. Note that,
when $V(t)\to\infty$, the solutions to this equation become the usual
ones for perfect reflectors, $k_m=mL_x/2$, with $m$ a positive integer.

Let us define
\begin{equation}
\Psi_{\bf m}({\bf x},t) = \left\{   \begin{array}{ll}
                 \psi_{\bf m}(x,y,z,t) & 0 \le x  \le L_x/2 \\
                 -\psi_{\bf m}(x-L_x,y,z,t) & L_x/2 \le x \le L_x
                      \end{array}
        \right.
\end{equation}
These functions satisfy the b.c. and the orthogonality relations
$$\left(\Psi_{\bf m}, \Psi_{\bf n}
\right)  = \left[ 1 - \sin(k_m(t) L_x) / k_m(t) L_x \right]
\delta_{\bf m,n}\, .$$ 
There is a second set of modes with a node on the cavity midpoint.
As these solutions do not ``see" the slab, they will be irrelevant in what follows.

For $t\leq 0$ the slab is not irradiated, consequently $V$ is
independent of time and has the value $V_0$. The modes of the
quantum scalar field that satisfy the Klein Gordon equation
(\ref{fieldequation}) are
\begin{equation}
u_{\bf m}({\bf x},t)=\frac{e^{-i \bar \omega_{\bf m} t}}{ \sqrt{
2\bar\omega_{\bf m}}}\Psi_{\bf m}({\bf x},0)\,\,\, , \label{u}
\end{equation}
where  $\bar\omega_{\bf
m}^2=(k_m^0)^2 + \left(\frac{\pi m_y}{L_y}\right)^2 +
\left(\frac{\pi m_z}{L_z}\right)^2$ and $k_m^0$ is the $m$-th
solution to Eq.(\ref{trascendentalequation}) for $V=V_0$. At
$t=0$ the potential starts to change in time and the set of 
numbers $\{ k_m
\}$  acquires a time
dependence through Eq.(\ref{trascendentalequation}).

Using Eq. (\ref{u}) we expand the field operator $\phi$ as
\begin{equation}
\phi({\bf x},t) =
\sum_{\bf m} \left[ b_{\bf m} u_{\bf m}({\bf x}, t)
+ b^\dagger_{\bf m} u^*_{\bf m}({\bf x}, t) \right],
\end{equation}
where $b_{\bf m}$ are annihilation operators.
Notice that in the above equation  we omitted the modes
with a node at $x=L_x/2$ because  their dynamics is not 
affected by the presence of the slab.

For $t\geq 0$ we write the expansion of the field mode $u_{\bf s}$ as
\begin{equation}
u_{\bf s}({\bf x},t>0) = \sum_{\bf m} P_{\bf m}^{({\bf
s})}(t) \Psi_{\bf m}({\bf x},t) .
\label{singlefunction}
\end{equation}
Assume a time dependent
conductivity given by
\begin{equation}
V(t) = V_0 + \left(V_{\rm max} - V_0\right) f(t)\,\,\, ,
\label{vgen}
\end{equation}
where $f(t)$ is a periodic and non-negative function,
$f(t)=f(t+T)\geq 0$, that vanishes at $t=0$ and attains its maximum
at $f(\tau_e)=1$. In each period, $f(t)$ describes the excitation
and relaxation of the plasma sheet produced by the laser pulse.
Typically, the characteristic time of excitation $\tau_e$ is the
smallest time scale and satisfies $\tau_e\ll T$. 
Under certain constraints, large changes in $V$  induce only small
variations in $k$ through the transcendental relation between $k$
and $V$ (see Eq. (\ref{trascendentalequation})). In this case, a
perturbative treatment is valid and a linearization of such
a relation is appropriate. Accordingly we write
\begin{equation} k_n(t)=k_n^0
(1+\epsilon_n f(t)) , \label{kdet} 
\end{equation} where 
 \begin{equation} \epsilon_n = \frac{V_{\rm
max}-V_0}{L_x (k_n^0)^2 + V_0 \left(1+\frac{V_0 L_x}{4}\right)}
.\label{epsilon} \end{equation} The restriction for the validity of the
perturbative treatment is $V_0 L_x \gg V_{\rm max} / V_0>1$. These
conditions are satisfied for realistic values of $L_x$, $V_0$, and
$V_{\rm max}$. 

Replacing  Eq.(\ref{singlefunction})
into $(\nabla^2 -
\partial^2_t) u_{\bf s}=0$  we find
a set of coupled differential equations for the amplitudes $P_{\bf m}^{({\bf
s})}(t) $. The dynamics is described  by  a set of coupled
harmonic oscillators with periodic frequencies and couplings,
as already discussed in this section.  It
is of the same form as the equations that describe the modes of a
scalar field in a three dimensional cavity with an oscillating
boundary. For the same reasons as before, a naive perturbative solution of previous equations
in powers of $\epsilon_n$ breaks down after a short amount of time  when the 
external frequency is tuned with some of the
eigenfrequencies of the cavity. 
Assuming that $f(t)$ is a sum of harmonic functions of frequencies
$\Omega_j=j2\pi/T$, the resonance condition is
$\Omega_j=2\tilde{\omega}_{\bf
n}$ for some $j$ and ${\bf n}$. If there is no intermode coupling, a nonperturbative solution 
gives an exponential
number of created photons  in that particular mode 
\begin{equation}
\langle {\mathcal N}_{\bf n}(t) \rangle = \langle b^\dagger_{\bf n} b_{\bf n}\rangle
=\sum_{\bf s} 2 \bar\omega_{\bf n} |A_{\bf n}^{({\bf s})} (t)|^2 \approx
\sinh^2 \left(\frac{(k^0_n)^2 f_j}{\Omega_j}  \epsilon_n t \right),
\label{photonnumber}
\end{equation}
where $f_j$ is the amplitude of the oscillations of $f(t)$ with frequency $\Omega_j$.

A full electromagnetic calculation has been presented in Ref.\cite{jap}.
It was shown there that the scalar model presented here describes the TE electromagnetic modes
inside the cavity. The treatment of TM modes involves
an independent scalar field,
with a potential proportional to $\delta'(x-L_x/2)$. 
Moreover, the model has also been generalized to the case of
arbitrary positions of the plasma sheet within the cavity \cite{jap}. 
The number of created TE photons depends strongly on the position 
of the layer, and  the maximum number is attained when it is located at the midpoint of the cavity. On the other hand, for TM modes this dependence is rather weak. 

In the treatment above no dissipation effects are considered (the delta-potential is real).
Similar calculations for lossless dielectric slabs with time-dependent and real permittivities \cite{schutz}
also neglect dissipation. However, it has been pointed out that dissipative effects may be relevant
in the evaluation of created photons \cite{Dodonov06}. In general, one expects the electromagnetic energy to be dissipated in the cavity walls,  in the plasma sheet, and/or 
in dielectric slabs contained in the cavity. In resonant situations without  dissipation, we
have seen that the dynamics of the relevant electromagnetic mode is described by
a harmonic oscillator with time dependent frequency.   A phenomenological way of taking into account
dissipative effects is to replace this equation with that of a damped oscillator
\cite{dodonov09} . Of course
this model cannot be consistently quantized unless one includes a noise term, 
otherwise the usual commutation relations are violated. Using the {\it quantum noise operator 
approach} \cite{noise} one can estimate the rate of photon creation in this model and,
provided the dissipation is not too large, the number of photons still grows exponentially,
although at a smaller rate. 
However, it has been recently argued  \cite{Dezael10} that these 
results should be valid only in the short time limit, while in the long time limit the system should
reach a stationary state with a constant number of photons inside the cavity. As the calculations in  \cite{Dezael10} 
involve 1D cavities, this point deserves further investigation. 

\section{Experimental perspectives}

Since the first theoretical predictions about motion induced
radiation, it was clear that the experimental observation
of this effect was not an easy task. As mentioned at the end of Section 2, the
photon creation produced by
a single accelerated mirror is extremely small in realistic
situations (see Eq. (\ref{NT3D})). 

The most promising situation seems to be the photon creation by
parametric amplification described
in Section 4.  However simple numerical estimations show that, even in the
most favorable cases,
it is difficult to observe the DCE in the laboratory.
In all the 3D examples discussed in Section 4, the number of created photons
grows exponentially in time as
 \begin{equation}
\langle {\mathcal{N}} \rangle =   \sinh^{2}\left(\eta\omega\epsilon t\right)\, ,
\label{exp growth}
\end{equation}
where $\omega$ is the frequency of the resonant mode and $\eta$ is a number of order $1$
related to the geometry of the cavity. Here $\epsilon$ denotes the relative 
amplitude of the oscillations in the moving mirror case,  or the relative amplitude 
of the oscillations of the relevant component of the  wavevector in the case of time dependent conductivity (see Eq.(\ref{kdet})).    
This equation is valid as long as $\epsilon^2\omega t\ll 1$ and
neglects any dissipative effects. As the electromagnetic cavity has a finite $Q$-factor, a rough estimation
of the maximum number of created photons $\langle {\mathcal{N}}_{max} \rangle $
is obtained by setting $t_{max}= Q/\omega$ in the above equation. As
mentioned at the end of Section 4, there is no  agreement
in the literature about this estimation. Calculations based on the use
of a master equation \cite{Dodonov98} give an exponential
growth with a rate diminished by a factor $\Gamma=1-1/(2Q\epsilon)$ (see also Ref. \cite{Mendonca}). 
On the other hand, it was shown that, in the case of 1D cavities, the 
total number of photons inside the cavity should reach a constant value
proportional to the finesse of the cavity at
long times \cite{Dezael10}. It was argued in the same work that
the exponential growth in the presence of dissipation would be valid only at short times.
 In any case, it is clear
that a {\it necessary} condition to have an observable number of photons is
that $2Q\epsilon>1$.

Assuming a cavity of length $L_0\simeq 1$ cm, the oscillation frequency 
of the mirror should be in the GHz range in order to meet the parametric resonance condition. 
A plausible possibility for reaching such high mechanical oscillation
frequencies is to consider surface vibrations, instead of
a global motion of the mirror \cite{Dodonov96}. In this context,
the maximum attainable values of the relative amplitude would be  
around $\epsilon\simeq 10^{-8}$, and therefore
the quality factor of the cavity should be greater than $10^8$  in order
to have a non-negligible number of photons. Microwave superconducting cavities with Q-factors as high as $10^{12}$ have been built \cite{Arbert-Engels01}. 
However, the Q-factor would be severely limited by the presence of an oscillating wall. 
Therefore, it is an extraordinary challenge to produce
extremely fast oscillations  while keeping the
extremely high  $Q$-factors needed in the DCE. Moreover,
the oscillations should be tuned with high precision to parametric resonance
with a cavity mode. 

\subsection {High frequency resonators and photon detection via
superradiance}

A concrete setup for producing and detecting
motion induced photons has been proposed in Ref.\cite{Kim06}.
A Film Bulk Acoustic Resonator (FBAR) is a device that consists of 
a piezoelectric film sandwiched between two electrodes.
An aluminum nitride FBAR of thickness corresponding to a half of the 
acoustic wavelength can be made to vibrate up to
a frequency of $3$GHz, with an amplitude of $\epsilon=10^{-8}$. The
expected maximum power
of Casimir photons produced by such  a FBAR
depends of course on the $Q$-factor of the cavity. It can be estimated to be
\begin{equation}
P_{max}=\langle {\cal N}_{max}\rangle \hbar\omega/t_{max}\, .
\end{equation}
Assuming that $Q\epsilon
= O(1)$, this gives $P_{max}\simeq 10^{-22}W$, which is too small
for direct detection. 

However, this low power could be detected using
ultracold atoms.
Let us consider a cavity  filled with an ensemble of population-inverted atoms
in a hyperfine state whose transition frequency is equal to the resonance
frequency of the cavity.
Then the Casimir photons can trigger a stimulated emission of the
atoms, and therefore they can
be indirectly detected by this form of {\it superradiance}.
Ref.\cite{Kim06} contains
a description of the experimental setup that could be used
to observe the DCE, and a
discussion about the rejection of signals
not produced by the Casimir photons. In particular, stimulated
amplification could also be triggered by the spontaneous
decay of one of the atoms (superfluorescence). In order to discriminate between
both effects it could be necessary to attain larger values of $Q\epsilon$.

\subsection {Time dependent conductivity induced by ultra-short laser pulses}

In order to avoid the experimental complications associated
with the high frequency motion of the mirror, it is possible to produce
effective changes
in the length of the cavity by inducing abrupt changes in the
reflectivity of a slab
contained in the cavity, as already mentioned in Section 4.3. This can be done by illuminating a
semiconducting slab with ultra short laser pulses \cite{Yablonovitch89,Lozovik95}.
An experiment based on this idea is currently being carried out
by the group of Padova \cite{Padova08}.

In this case, a numerical estimation of the maximum number of photons
created in the cavity
looks, at first sight, much more promising than in the case of moving mirrors.
Using a slab of a thickness around $1$ mm, it is possible to reach
values of $\epsilon$ as large as $10^{-4}$, and therefore the constraints on the
$Q$-factor of the cavity are considerably milder. Moreover, 
it is experimentally possible to generate trains of thousands of laser pulses with
a repetition frequency on the order of a GHz.

In Padova's setup (see Fig. 4), a high $Q\approx 10^6$
superconducting cavity contains a GaAs
semiconducting slab. The laser pulses are tuned at $4.70$ GHz, twice
the frequency of the fundamental
TE mode of the cavity. However, 
it has been pointed out \cite{Dodonov06} that dissipative effects may
play an important role in this kind of experiment.
Indeed, the changes in the conductivity of the slab
are due to the creation of 
electron-hole pairs by the laser pulses, and during this process
the dielectric permittivity acquires an imaginary part. The associated
dissipation prevents photon creation unless severe
constraints on the properties of the semiconductor are fulfilled: it must have a
very short recombination time (tenths of
ps), and a high mobility (around $1 m^2 (Vs)^{-1}$). A slab with these characteristics has been
constructed by irradiating a GaAs sample  with fast neutrons, in order to 
reduce the recombination time of the original sample (about $1$ ns) while keeping constant the value of the 
mobility \cite{Padova08}.
Photons are detected using a loop antenna
inside the cavity. The minimum number of photons that can be detected is around $100$,
below the expected signal of Casimir photons \cite{Braggio09,Padova08}.

A related setup is  illumination of a {\it superconductor} instead of a semiconductor surface. 
The advantage in this case is that dissipative effects are less important, because the variation
of the imaginary part of the permittivity is much smaller for superconductors than for semiconductors
in the microwave region \cite{Segev07}. Moreover, since the abrupt changes in the conductivity are due to local heating
of the surface (and not to the creation of electron-hole pairs as in the semiconductor), the intensity of the laser can be
considerably smaller, reducing unwanted effects of energy accumulation inside the cavity.

\begin{figure}[b]
\begin{center}
\includegraphics[scale=0.2]{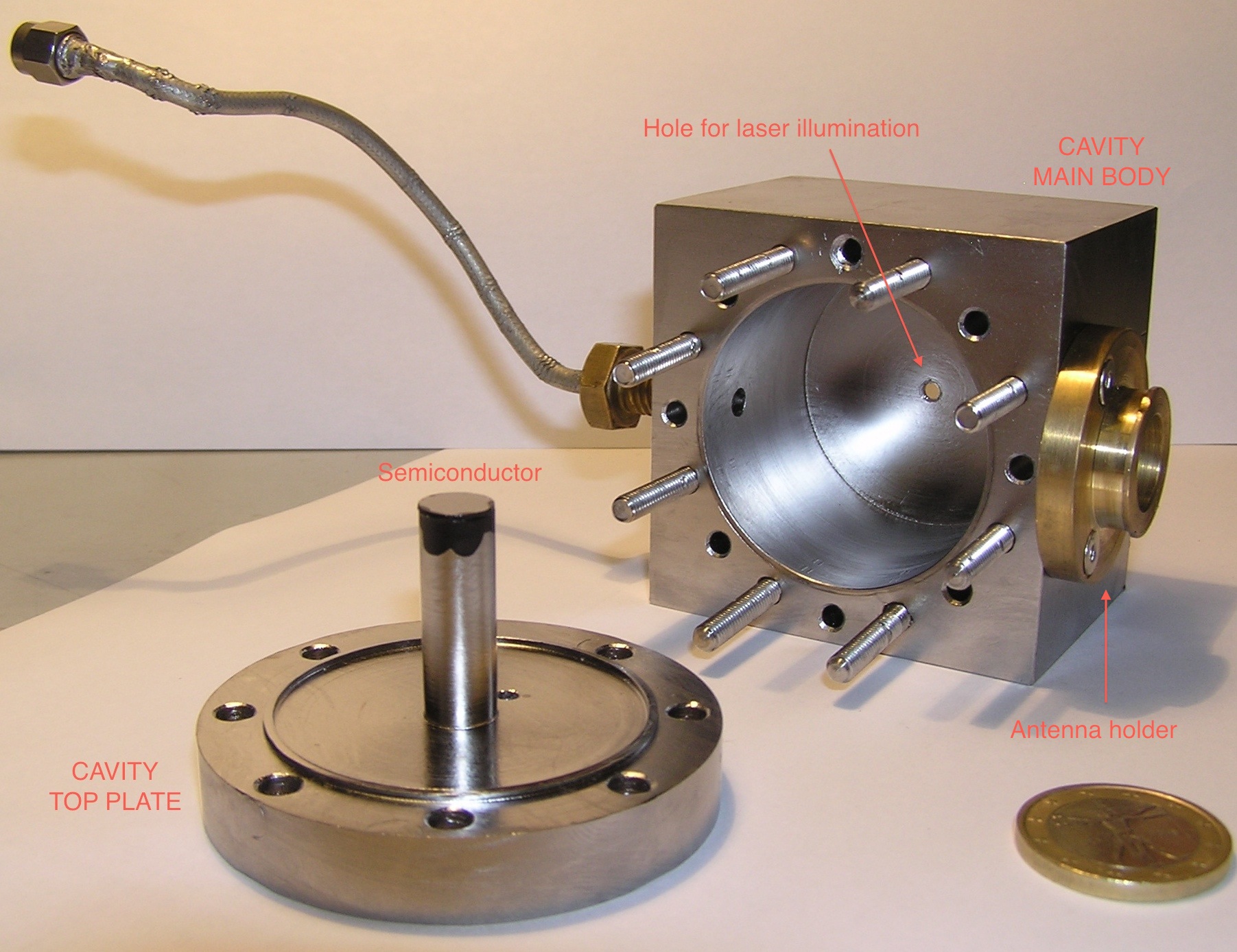}
\end{center}
\caption{ Superconducting cavity of the experimental setup of MIR experiment at Padova to measure the dynamical Casimir effect (Courtesy of Giuseppe Ruoso).}
\label{density}
\end{figure}

\subsection{Optical parametric oscillators}

Standard nonlinear optics can be interpreted, in some cases, as a time-dependent modulation of 
the refractive index. 
Ref.~ \cite{Dezael10} considered 
an optical parametric oscillator (OPO) with a pump laser beam of frequency $\Omega$ and amplitude $E_{\rm pump}$ interacting with 
a very thin  $\chi^{(2)}$  nonlinear crystal slab placed on the interior side of the cavity mirror.   For a type-I arrangement, 
the total polarization component along a suitable crystal symmetry axis  may be written in terms of the intracavity electric field component along the same direction as  \cite{Dezael10}
\begin{equation}
\label{nonlinearP}
P(t) =  \epsilon_0 \left( \chi^{(1)} + \frac{1}{2} \chi^{(2)} E_{\rm pump} \sin(\Omega t -\theta) \right) E(t) ,
\end{equation}
where $\epsilon_0$ is the vacuum permittivity,
$\chi^{(1)}$ and $\chi^{(2)}$ are the relevant components of the linear and second-order nonlinear susceptibility tensors,
and $\theta$ is a phase that depends on the pump beam phase and the position of the crystal. 

The total susceptibility 
as given by eq. (\ref{nonlinearP}) (including the nonlinear second-order term) 
corresponds to an effective refractive index oscillating at the pump beam frequency $\Omega,$ 
thus leading to a 
modulation of the optical cavity length.
This is formally equivalent to modulating the physical cavity length by bouncing the mirror with frequency $\Omega.$
But in the OPO the pump beam frequency is in the optical range, as are the generated photons with frequencies satisfying 
$\omega+\omega'=\Omega.$ For $\omega$ and $\omega'$ corresponding to cavity modes, parametric amplification is enhanced and the resulting 
photon flux is typically several orders of magnitude higher than in the case of mechanical motion \cite{Dezael10}.

\subsection{Superconducting coplanar waveguides}

Another possibility to induce fast variations of the b.c. on the electromagnetic field is to consider a 
coplanar waveguide terminated by a SQUID  \cite{Nori09,Delsing10}.
A time-dependent magnetic flux can be applied  to control the effective inductance of the SQUID, which in turn produces a time-dependent 
Robin boundary condition for the  phase field (time integral of the electric field), equivalent to that of a transmission line
with a variable length. This setup simulates 
a moving Robin mirror in 1D  with an effective velocity that might be close to the speed of light. As a consequence, 
 the first-order non-relativistic results \cite{Lambrecht96,Mintz06B}, based on the perturbative approach outlined in Section 2,  must be modified by
the inclusion of higher-order frequency sidebands  \cite{Nori09}. 

As in the previous examples, it is crucial to check if the flux of Casimir photons can be discriminated
from other sources of photons, like the classical thermal contribution. The analysis presented in Ref.\cite{Nori09}
shows that this is the case, for realistic values of the parameters, at temperatures below $70$ mK. 

\section{Final remarks}

We have reviewed some theoretical and experimental advances 
in the analysis of moving bodies or time dependent boundary
conditions coupled
to the vacuum fluctuations of the electromagnetic field.

Accelerated neutral bodies produce the emission of real photons, while experiencing a radiation reaction force.  When the dynamics of the bodies 
is treated quantum mechanically, 
the interaction with the vacuum fluctuations
not only causes this dissipative force,  but also an appreciable amount of decoherence,
which is a consequence of the entanglement between the mirrors and the electromagnetic 
field. This is a particular example of quantum Brownian motion, where the 
Brownian particle (mirror) loses coherence while being subjected to a damping force
due to its coupling to the environment (the quantum field). 

When two neutral bodies are in relative motion,
we expect velocity-dependent forces between them. There is a particularly
interesting situation in which two parallel, non-perfectly 
conducting slabs 
are in relative parallel motion with constant velocity. In this case,
there is a {\it vacuum friction} between the slabs even in the absence of
real photons. The effect can be understood in terms of the interaction of 
image charges, or as the interchange of virtual photons between the surfaces.
There are similar friction forces for neutral atoms moving near surfaces.

The rate of photon creation produced by a single accelerated body in
free space is deceptively small
in realistic situations. However, in closed cavities a much larger
number of photons may be produced
by parametric amplification. Indeed in an ideal 3D cavity one expects
an exponential growth in the number of photons
when its size varies periodically at an appropriate resonant
frequency, making detection of photon creation not an impossible
task.  The calculation of  photon creation in the presence of
ideal conductors have been performed in 1D and 3D using different analytical 
approximations. The results are consistent and have been confirmed 
by fully numerical calculations. However, the case of moving mirrors with finite conductivity 
(i.e electromagnetic cavities with a finite $Q$-factor)
is not a completely settled issue. In 1D cavities, at short times the growth of the total number of
created photons is
still exponential (with a different rate), while at large times the total 
number of photons should reach saturation.
This problem has not yet  been solved  for 3D cavities. The difficulties in evaluating
the DCE for mirrors with finite conductivity recalls a similar 
situation  in the static Casimir effect, where the evaluation of the Casimir force depends
strongly on the theoretical model used to describe the conductivity of the 
bodies,  and there are interesting correlations between finite conductivity, temperature
and geometry. These correlations may have relevant counterparts in the dynamical problem.
In any case, although difficult, the direct
experimental detection of the motion induced radiation is not out
of reach, as long as one can keep a very high $Q$-factor in 
a  cavity with moving walls.  In particular,
there is a specific proposal that involves nanoresonators in
a high Q-cavity filled with a gas of cold atoms
to detect a small number of photons through superradiance.

An exponentially large number of photons can
also be produced when some electromagnetic property of
the cavity varies periodically with time. Of particular importance is
the case in which  the conductivity of a semiconductor
or superconductor slab placed inside an electromagnetic cavity is modulated using
short laser pulses. Theoretical estimates show that
this setup could be implemented, with milder requirements on the
$Q$-factor of the cavity. Once again,
there is no comprehensive theoretical model that takes  into
account the (dissipative) response of the slab to the laser pulses,
and  its relevance for the photon creation
process. However, this is a promising
alternative and there is an ongoing experiment at 
Padova  based on this setup.

There are other possibilities  to produce fast variations of the boundary conditions
on the electromagnetic field, that involve optical parametric oscillators or 
superconducting waveguides. The theoretical analyses  suggest
that it should be easier to detect the photons created in these setups than in the
case of  moving mirrors.

In summary, there is a plethora of interesting effects related to the electromagnetic vacuum fluctuations
in the presence of moving bodies and/or other time dependent external conditions. The eventual 
experimental confirmation of some of these effects will certainly produce an increasing activity
on this subject in the near future, as was the case for the static Casimir effect following the first realization
of precise experiments in that area since 1997.

\bigskip

\noindent{\bf Acknowledgements} The work of DARD was funded by DARPA/MTO's Casimir Effect Enhancement program under DOE/NNSA
Contract DE-AC52-06NA25396. PAMN thanks CNPq and CNE/FAPERJ
for financial support and the Universidad de Buenos Aires
for  its hospitality during his stay at Buenos Aires. FDM thanks the Universidad de Buenos Aires, 
CONICET and ANPCyT for financial support.  We are grateful to Giuseppe Ruoso for providing a picture of MIR experiment at Padova.

\end{document}